\documentclass[11pt]{article}

\newsavebox{\junk}
\savebox{\junk}[1.6mm]{\hbox{$|\!|\!|$}}

\def\bbbz{{\mathchoice {\hbox{$\sf\textstyle Z\kern-0.4em Z$}}
{\hbox{$\sf\textstyle Z\kern-0.4em Z$}}
{\hbox{$\sf\scriptstyle Z\kern-0.3em Z$}}
{\hbox{$\sf\scriptscriptstyle Z\kern-0.2em Z$}}}}
\def\sq{\hbox{\rlap{$\sqcap$}$\sqcup$}}

\usepackage{amsmath} 
\usepackage{amsfonts}
\usepackage{latexsym}
\usepackage{bm}

\newcommand{\ben}{\begin{enumerate}}
\newcommand{\een}{\end{enumerate}}
\newcommand{\bit}{\begin{itemize}}
\newcommand{\eit}{\end{itemize}}

\newtheorem{theorem}{Theorem}[section]
\newtheorem{proposition}[theorem]{Proposition}

\newcommand{\ba}{\begin{array}{rcl}}
\newcommand{\ea}{\end{array}}
\newcommand{\bt}{\begin{theorem}}
\newcommand{\et}{\end{theorem}}
\newcommand{\bd}{\begin{description}}
\newcommand{\ed}{\end{description}}

\def\slabel#1{\label{s:#1}}

\def\eq#1/{(\ref{e:#1})}

\def\Section#1{Section~\ref{s:#1}}

\def\beq{\begin{equation}}
\def\eeq{\end{equation}}
\def\beqa{\begin{eqnarray}}
\def\eeqa{\end{eqnarray}}

\def\qed{\ifmmode\sq\else{\unskip\nobreak\hfil
\penalty50\hskip1em\null\nobreak\hfil\sq
\parfillskip=0pt\finalhyphendemerits=0\endgraf}\fi}

\def\sqr#1#2{{\vcenter{\hrule height.#2pt
      \hbox{\vrule width.#2pt height#1pt \kern#1pt
         \vrule width.#2pt}
       \hrule height.#2pt}}}

\newcommand{\bc}{\begin{corollary}}
\newcommand{\ec}{\end{corollary}}

\newcommand{\bp}{\begin{proposition}}
\newcommand{\ep}{\end{proposition}}

\def\eye(#1){{\bf (#1)}\quad}

\def\taboo#1{{{}_{#1}}}

\def\0P{\taboo{0}P}
\def\0Pn{\taboo{0}P^n}

\usepackage{bbm}
\usepackage{amsmath}
\DeclareSymbolFont{largesymbols}{OMX}{yhex}{m}{n}
\DeclareMathAccent{\widewidehat}{\mathord}{largesymbols}{"62}

\usepackage[ruled,vlined]{algorithm2e}
\usepackage{color}
\usepackage{graphicx}
\usepackage[position=t,singlelinecheck=off]{subfig}
\usepackage{caption}
\usepackage{tikz}
\usepackage[version=4]{mhchem}
\usepackage{amsmath,amssymb,tikz-cd}
\usepackage[titletoc]{appendix}

\DeclareMathAlphabet{\mathpzc}{OT1}{pzc}{m}{it}

\date{May 6, 2021}

\usepackage{mathtools}
\usepackage{extarrows}
\usetikzlibrary{matrix}
\usepackage{multicol}

\usepackage{dutchcal}

\usepackage{tikz}
\usepackage{xcolor}
\usetikzlibrary{arrows}
\usetikzlibrary{decorations.markings}
\usetikzlibrary{positioning}
\usetikzlibrary{matrix,fit}
\pgfkeys{tikz/mymatrix/.style={matrix of math nodes,inner sep=0pt,row sep=0em,column sep=0em,nodes={inner sep=6pt}}}

\usepackage[format=plain,
            labelfont={bf,it},
            textfont=it,
            margin=0.05in,
            font=small]{caption}

\newcommand{\ds}{\displaystyle}

\newcommand{\genR}{$R_{i}$ }

\newenvironment{example}[2][Claim]{\begin{trivlist}
\item[\hskip \labelsep {\bfseries #1}\hskip \labelsep {\bfseries #2:}]}{\end{trivlist}}

\newenvironment{prop}[2][Proposition]{\begin{trivlist}
\item[\hskip \labelsep {\bfseries #1}\hskip \labelsep {\bfseries #2:}]}{\end{trivlist}}

\SetKw{KwDownTo}{down to}

\title{Coupling from the Past for the Stochastic Simulation of Chemical Reaction Networks}

\def\jem{Postal Address:
      Department of Applied Mathematics,
      University of Colorado, Box 526
      Boulder CO 80309-0526, USA; email: corcoran@colorado.edu, 
      phone: 303-492-0685}

\author {J.N. Mueller, J.N. Corcoran\\
University of Colorado \thanks{\jem} }

\begin{document}


\maketitle
\vspace{-.8cm}

\begin{abstract} \small \noindent

Chemical reaction networks (CRNs) are fundamental computational models used to study the behavior of chemical reactions in well-mixed solutions. They have been used extensively to model a broad range of biological systems, and are primarily used when the more traditional model of deterministic continuous mass action kinetics is invalid due to small molecular counts.
We present a \emph{perfect sampling} algorithm to draw error-free samples from the stationary distributions of stochastic models for coupled, linear chemical reaction networks. The state spaces of such networks are given by all permissible combinations of molecular counts for each chemical species, and thereby grow exponentially with the numbers of species in the network. To avoid simulations involving large numbers of states, we propose a subset of chemical species such that coupling of paths started from these states guarantee coupling of paths started from all states in the state space and we show for the well-known Reversible Michaelis-Menten model that the subset does in fact guarantee perfect draws from the stationary distribution of interest. We compare solutions computed in two ways with this algorithm to those found analytically using the chemical master equation and we compare the distribution of coupling times for the two simulation approaches.


\bigskip

\end{abstract}

\footnotetext{Keywords: chemical reaction networks, perfect sampling,
stochastic chemical kinetics, continuous-time Markov chains, birth-and-death processes \\
AMS Subject classification: 60J27,  60J28, 60K30}

\setcounter{page}{0}





\section{Introduction}
\slabel{int} 
A chemical reaction network (CRN) is a collection of chemical species and the rules governing interactions between them.
One of the most well known simple examples is given by the reversible Michaelis-Menten (RMM) model for an enzymatic reaction which is depicted as follows.

\tikzstyle{block} = [draw,draw=white,fill=white!20,minimum size=2em]
\begin{center}
\begin{tikzpicture}[auto, node distance=1cm,>=latex']
    \node [block, name=C] (C) {\mbox{ \large $C$} };
    \node [block, left of=C, node distance=2cm] (SE) {\mbox{ \large$S+E$} };
    \node [block, right of=C,node distance =2cm] (PE) {\mbox{ \large$P+E$} };
    
    \draw[thick,->] ([yshift= 0.08cm]SE.east) to [out=0,in=180] node[auto] {$k_{1}$} ([yshift= 0.08cm]C.west);
    \draw[thick,->] ([yshift= -0.08cm]C.west) to [out=180,in=0] node[auto] {$k_{2}$} ([yshift= -0.08cm]SE.east);
    \draw[thick,->] ([yshift= 0.08cm]C.east) to [out=0,in=180] node[auto] {$k_{3}$} ([yshift= 0.08cm]PE.west);
    \draw[thick,->] ([yshift= -0.08cm]PE.west) to [out=180,in=0] node[auto] {$k_{4}$} ([yshift= -0.08cm]C.east);        
\end{tikzpicture}
\end{center}

Here, {\ce{E}} represents an enzyme, {\ce{S}} a substrate, {\ce{C}} an enzyme-substrate complex, and {\ce{P}} represents a product. $k_{1}, k_{2}, k_{3}$ and $k_{4}$ are multipliers for reaction rates which are assumed to be proportional to the amount of each species. For example, the rate of formation of the complex {\ce{C}} from existing enzyme {\ce{E}} and substrate {\ce{S}} is $k_{1} es$ where $e$ and $s$ denote the concentrations of {\ce{E}} and {\ce{S}}, respectively. 


A common approach to modelling chemical reaction networks involves representing the time-rates-of-change in species' concentrations with a set of coupled differential equations known as the \textit{reaction rate equations}. Underpinning these models are the assumptions that the molecular concentrations vary continuously over time and evolve in a deterministic way \cite{meng2017}. Many homogeneous chemical systems, particularly reaction systems involving large numbers of molecules, can be well approximated by such models \cite{mcquarrie1967}. However, the processes underlying chemical reactions are inherently statistical in nature and the assumption that a reaction network can be represented as a continuous process may be invalid for some reactions such as those involving low molecular counts 
\cite{meng2017, gillespie1977, anderson2010, gillespie2001, melykuti2014}. Such networks should be modelled stochastically.

The stochastic modelling of a chemical reaction network, dating back to Delbr{\"u}ck \cite{delbruck}, is based upon the idea that molecular concentrations are subject to statistical fluctuation. Rather than treat the molecular counts as continuous quantities that evolve deterministically in time, the stochastic representation follows the time-evolution of the probability distributions of discrete states of a stochastic process \cite{meng2017, gillespie1977} where the collections of states is the set of all permissible combinations of molecular counts of the chemical species in the reaction network at a time $t \geq 0$ \cite{anderson2010, gillespie2007, levien2017}.

For simple CRNs the chemical master (Forward Kolmogorov) equation can be used to model the probability of having $n$ particles of a given species $i$ at a time $t$ \cite{delbruck}.
\begin{align}
    \frac{dP(X,t)}{dt} = \sum_{i=1}^n \big( \lambda_i(X - R_i) P(X - R_i,t) - \lambda_i(X) P(X,t)  \big)
    \label{e:CME}
\end{align}
Here $\lambda_i(X - R_i)$ and $\lambda_i(X)$ are the transition rates into and out of state $i$, respectively, and $P(X - R_i, t)$ and $P(X, t)$ are the probabilities of these transitions \cite{melykuti2014, gillespie2007}.

While the the chemical master equation deals more carefully with the probabilistic nature of dilute reaction mixtures, analytic solutions to \eqref{e:CME} are extremely difficult to compute and are known for only a handful of special cases \cite{meng2017, hertzberg1980, jahnke2006}. Alternate methods for computing analytical solutions, such as the novel ``gluing" methods introduced by \cite{melykuti2014} and explored in \cite{meng2017, melykuti2015}, have been proposed. Such solution methods decompose the state space into subsets whose solutions are straightforward to compute and then build up the full analytical solution by recursively ``gluing" solutions from each subset together at one or two states. Other studies have been made to characterize the \emph{form} of analytic solutions that arise for certain classes of chemical reactions networks \cite{anderson2018}. The \emph{product form solution} is one such example that is guaranteed to exist for closed, irreducible subsets of the state space of a stochastically modelled CRN provided that the network is weakly reversible and has deficiency zero \cite{anderson2018, feinberg89}

Commonly, numerical methods are required to find solutions to \eqref{e:CME} for most problems of interest. The simplest such models treat the CRN as a continuous time, discrete state Markov process. The state transitions are given by the various reactions at times $t$ modelled by a Poisson process which, among other things, assumes that the times until the next reaction are modelled as exponential random variables.

We are often interested in the time evolution of a system, started with known initial concentrations, at some later fixed time $t$. In this case, direct simulation methods such as the \textit{Stochastic Simulation Algorithm} \cite{gillespie2007} can be used to follow every reaction, which is obviously inefficient. Many indirect methods have been developed, most notably Gillespie's ``$\tau$-leaping'' method {\cite{gillespie2001}}, which take advantage of the rich properties of the Poisson process in order to simulate over finite time scales through much larger time steps. In the case when one is concerned with the stationary (equilibrium) behavior of the chemical reaction network, simulations are typically run for ``a long time'' until it is believed that convergence has, at least approximately, been achieved. In this paper, we derive ``perfect simulation'' algorithms which avoid convergence issues and enable us to sample values directly from the stationary distribution of the network.

The remaining sections of this paper are organised as follows: in \Section{B&D} we give a brief overview of birth-and-death processes and we describe non-perfect approaches for sampling from their stationary distributions. We then describe \emph{perfect} sampling and we provide a perfect sampling algorithm for birth-and-death processes. Using this algorithm, we compute the stationary distribution of a simple CRN that may be modelled as a birth-and-death process and we compare to the exact solution. In \Section{networks}, we extend the perfect sampling approach to more complicated CRNs and we propose a subset of states which may be used to compute the distribution of interest more efficiently. Finally, we relax certain assumptions made on the network and we provide numerical evidence that our approach for computing the stationary distributions with the proposed subset applies to larger and more complex chemical reaction networks.

\section{Overview and Simulation for Birth-and-Death Processes}
\slabel{B&D}

\subsection{An Overview of Birth-and-Death Processes}
A {\emph{birth-and-death}} process is a Markov process $\{X(t)\}$ on a state space ${\mathbb{S}}=\{0,1,2,\ldots\}$ where $X(t)$ represents a population size at time $t$. $X(t)$ will increase and decrease over time due to events that can be thought of as ``births" and ``deaths" in the population. When the current population size is $i>0$, the next change in size will be either to state $i+1$  after an exponential amount of time with rate $\lambda_{i}$ or to state $i-1$ after an independent exponential amount of time with rate $\mu_{i}$. The parameters $\lambda_{i}$ and $\mu_{i}$ are known as birth and death rates for the process, respectively.

If the population size is currently $i$, the time until the next event occurs (a birth or death) is the minimum of two independent exponential times and thus has an exponential distribution with rate $\lambda_{i}+\mu_{i}$. This next event may be assigned to be a birth with probability $\lambda_{i}/(\lambda_{i}+\mu_{i})$ or a death with remaining probability $\mu_{i}/(\lambda_{i}+\mu_{i})$. The {\emph{stationary}} or {\emph{equilibrium}} distribution for this Markov process is given by
\beq
\label{e:bdstat}
\pi_{n} = \frac{\lambda_{0} \lambda_{1} \cdots \lambda_{n-1}}{\mu_{1} \mu_{2} \cdots \mu_{n}} \, \pi_{0}, \qquad \mbox{for} \,\, n \geq 1
\eeq
where $\pi_{n}$ represents the long-run probability of finding the process in state $n$. One would solve for $\pi_{0}$ by setting $\sum_{n=0}^{\infty} \pi_{n} = 1$. In order for a stationary distribution to exist, this sum must be convergent which means that there may be some restrictions on the birth and death rate parameters. We refer the interested reader to \cite{resnick} for more details about birth-and-death processes.

\subsection{Non-Perfect Simulation for Birth-and-Death Processes}
\slabel{nonperf}

Let $\{X(t)\}$ be a birth-and-death process on a finite state space $\mathbb{S} = \{0, 1, 2, \ldots, m\}$. Suppose that, at any instant when $X(t)=i$, the time until the next birth is exponentially distributed with rate $\lambda_{i}$ and the time until the next death is exponentially distributed with rate  $\mu_{i}$. We wish to sample from the stationary distribution of $\{X(t)\}$. 

With standard ``forward" simulation of $X(t)$, it is tempting to discretize time and consider simulating at each time step the next ``event" without regard to the amount of time that has passed. Indeed, we are looking to simulate long-run proportions of visits to various states and not the evolution of the system over a specific time interval. Unfortunately, a discrete time process embedded at the times of events will not have the same stationary distribution of our process of interest. We can, however, use the concept of {\emph{uniformization}} which involves running another process $\{Y(t)\}$ at the fastest possible, but notably constant, transition rates and then {\emph{thinning}} out the events in $\{Y(t)\}$ by considering each as a potential event in $\{X(t)\}$ with  appropriate probabilities \cite{serfozo1979, vandijk1992}. Specifically, if $\{X(t)\}$ is a birth-and-death process on ${\mathbb{S}}$ with birth rates $\{\lambda_{i}\}_{i \in {\mathbb{S}}}$ and death rates $\{\mu_{i}\}_{i \in {\mathbb{S}}}$, we define
$$
\lambda^{*} = \max_{i \in {\mathbb{S}}} \{\lambda_{i}\} \,\,\,\, \mbox{and} \,\,\,\, \mu^{*} = \max_{i \in {\mathbb{S}}} \{\mu_{i}\}.
$$
At any given moment, the time until the next event in $\{Y(t)\}$  is an exponential random variable with rate $R$. If an event is occurring in $\{Y(t)\}$, we assign it to be a birth with probability $\lambda^{*}/(\lambda^{*}+\mu^{*})$ and a death with probability $\mu^{*}/(\lambda^{*} + \mu^{*})$. Any birth (death) in $\{Y(t)\}$ has the potential to be a birth (death) in the process of interest $\{X(t)\}$. At any point in time where $\{Y(t)\}$ has a birth, we assign a birth in $\{X(t)\}$ at that time with probability $\lambda_{i}/\lambda^{*}$ where $i$ is the current state of the process $\{X(t)\}$. With probability $1-\lambda_{i}/\lambda^{*}$, we do nothing with  $\{X(t)\}$ at this time point. Similarly, at any point in time where $\{Y(t)\}$ has a death, we assign a death in $\{X(t)\}$ at that time with probability $\mu_{i}/\mu^{*}$ or do nothing in $\{X(t)\}$ with probability $1-\mu_{i}/\mu^{*}$. One can show that the time-embedded process that is $\{X(t)\}$, considered only at the event times for $\{Y(t)\}$, is a discrete time process $\{X_{n}\}$ whose stationary distribution is the same as that of $\{X(t)\}$ \cite{serfozo1979}.

\subsection{Perfect Simulation}
\slabel{perfect}

Most Markov chain Monte Carlo (MCMC) algorithms for sampling from stationary distributions involve simulating the model transitions for a ``long time" until the dynamics have settled down and convergence to the stationary distribution is approximately achieved. In this paper, we use a {\emph{perfect simulation}} (also known as {\emph{perfect sampling}} or {\emph{coupling-from the-past}}) approach \cite{prowil96} which will allow us to produce exact draws (samples) from the stationary distribution without any convergence issues.

The essential idea of perfect simulation is to find a random epoch in the past, denoted by 
$-T$ and described as a {\emph{backward coupling time}},  such that, 
if we construct sample paths from every possible value of $(X_{1}, X_{2}, \ldots X_{n})$ 
starting at time $-T$ and running forward,  all paths will have come together or ``coupled" by time zero. We may also refer to the positive value $T$, which gives the number of steps to start back in time rather than the starting time itself, as a backward coupling time. Either way, the common value of these paths at time zero is an exact draw from the stationary distribution.

Intuitively, it is clear why  this result holds with such a
random time $-T$. Consider a chain starting at $-\infty$ according to a draw from the  the stationary distribution. At every
iteration it necessarily maintains this distribution. At time $-T$ it must pick
{\it some} value $x$, and from then on it follows the trajectory from that
value. By construction of $T$, 
it arrives at
the same place at time zero no matter what value $x$ is picked at time $-T$,
so the value returned by the algorithm at time zero must itself be a draw 
from the stationary distribution. It is the tail end of a sample path that has run for an infinitely long time.

In an ordered space, perfect simulation algorithms can be particularly efficient 
if the chain is \emph{stochastically monotone} in the sense that paths from ``lower" 
starting points stay below paths from ``higher" starting points. In this case, one need
only couple sample paths from the ``top'' and ``bottom'' of the space, as all
other paths will be sandwiched in between. More generally, if no obvious ordering exists, 
it may still be possible to identify a 
subset of states such that coupling of sample paths started from all points in the 
subset implies coupling of sample paths started from all points in the state space.

There are several easy-to-read perfect sampling tutorials available and we
refer interested readers to \cite{casrob00}. In this 
paper we only wish to emphasize that the key idea in the search successively 
further and further back in time for the so-called \emph{backward coupling 
time} $T$ requires that one reuse random number streams. 
That is, if sample paths that run forward to time $0$ from time $-1$ using a 
random number (or random vector) $U_{-1}$ have not coalesced by time $0$, 
then one must go back further, say to time $-2$, and run paths forward for two
steps using a random number $U_{-2}$ and then the {\bf previously used} 
$U_{-1}$.

\subsection{Perfect Simulation for Birth-and-Death Processes}
\label{s:perfB&D}

We now describe a backward coupling algorithm for a birth and death process $\{X(t)\}$ on a finite state space ${\mathbb{S}} = \{0,1,2,\ldots, m \}$. Although a perfect simulation algorithm is not necessary here, as we can compute and draw values directly from the probability distribution given by (\ref{e:bdstat}), this construction will be useful for more complicated CRNs. 

It will be convenient for us to move from the standard notation of using $i$ to denote a state of the birth-and-death process to using $x$.

Consider starting sample paths from all of the $m+1$ possible states at some time $-n$. We will run paths forward by uniformization or ``thinning" a maximum rate process. A path at state $x$ will move to state $x+1$ with birth rate $\lambda_{x}$ and will move to state $x-1$ with death rate $\mu_{x}$. Denote the maximum birth and death rates over all possible paths as
$$
\lambda^{*} := \max_{x \in \mathbb{S}} \{ \lambda_{x} \} \,\,\,\, \mbox{and} \,\,\,\, \mu^{*}  := \max_{x \in \mathbb{S}} \{ \mu_{x} \}
$$
respectively. Let $R:= \lambda^{*} + \mu^{*}$. The paths, using the correct rates, that we are trying to couple is run conditionally based on events in the maximum rate process. Any birth in the maximum rate process is a potential birth for each individual population path and any death in the maximum rate process is a potential death for each individual path. The specific details for the perfect simulation algorithm are outlined in Algorithm \ref{alg:perfect_BD}. Note that we are moving backwards in blocks of length $N$ rather than backing up one step at a time. This is valid because if, for example, all paths started at time $-3$ are coupled by time $0$, then all paths that are started further back in time will end up at various locations at time $-3$ that will then be coupled together by time $0$. In other words, while we may miss the minimum backward coupling time, we will still find {\emph{a}} backward coupling time. We wish to stress again that each time we start over further back in time, the uniform variates at later time points on the journey to time $0$ must be reused.

\begin{algorithm}[H]
\caption{Perfect Birth-and-Death}
\begin{enumerate}
\item[0.] Choose a fixed integer $N \geq 1$. Set $n=N$.
\item Generate and store $2N$ independent  and identically distributed random variables $U_{-n}, U_{-n+1}, \ldots, U_{-n+N-1}$ and $V_{-n}, V_{-n+1}, \ldots, V_{-n+N-1}$ that are uniformly distributed over the interval $(0,1)$.
    
\item Start $m+1$ sample paths at time $-n$ by setting
$$
X_{i}^{(-n)} = i
$$
for $i=0,1,2,\ldots, m$.

\item For each time step $t=n,n-1,\ldots,1$, determine whether there is a birth or death in the maximum rate process $\{Y(t)\}$ as follows.
\begin{itemize}
\item If $U_{-t} \leq \lambda^{*}/(\lambda^{*} + \mu^{*})$, there is a birth in the maximum rate process. In this case, all paths have the potential for a birth.

Specifically, for $i=0,1,2,\ldots,m$, let $x = X_{i}^{(-t)}$ and assign
$$
X_{i}^{(-t+1)} = \min( x+1,m)
$$
whenever $V_{-t} \leq \lambda_{x}/\lambda^{*}$. Otherwise, assign $X_{i}^{(-t+1)}=x$.

\item If $U_{-t} > \lambda^{*}/(\lambda^{*} + \mu^{*})$, there is a death in the maximum rate process. In this case, all paths have the potential for  a death.

Specifically, for $i=0,1,2,\ldots,m$, let $x = X_{i}^{(-t)}$ and assign
$$
X_{i}^{(-t+1)} = \max(x-1,0)
$$
whenever $V_{-t} \leq \mu_{x}/\mu^{*}$. Otherwise, assign $X_{i}^{(-t+1)}=x$.
    
\end{itemize}
    
\item If $X_{0}^{(0)} = X_{1}^{(0)} = \cdots = X_{m}^{(0)}$, stop the algorithm. This common value is a perfect draw from the stationary distribution for the birth-and-death chain.

Otherwise, set n = n+N and return to Step 1.
    
\end{enumerate}
\label{alg:perfect_BD}
\end{algorithm}

\section{Modelling a Two-Species Network as a Birth-and-Death Process}
\slabel{networks}

We consider a chemical mixture consisting of $n$ different chemical species $S_{1}, S_{2}, \ldots, S_{n}$ in a well-stirred and fixed volume solution at a fixed temperature. We will assume that there are $X_{i}(t)$ molecules of species $i$ in the mixture at time $t$ and we will define $X(t) = (X_{1}(t), X_{2}(t), \ldots, X_{n}(t))$ to be the current state of the system. We suppose that the species combine to form complexes (for example $S_{1}+S_{2}$) and that  $m$ reversible reactions between the species are possible. We will label these reactions as $R_{1}, R_{2}, \ldots, R_{m}$. It is assumed that the system is subject to {\emph{mass action kinetics}} which implies that reaction rates are proportional to the amounts or concentrations of reacting species. In particular, there is a \emph{reaction rate constant} $k_{i}>0$ associated with reaction $R_{i}$. 

In this paper, we are interested in the joint distribution of chemical species
$X(t)=(X_{1}(t), X_{2}(t), \ldots, X_{n})(t)$ as it evolves over time as a Markov chain. Specifically, we are interested in this distribution for networks that reach an equilibrium in the Markov sense (as opposed to a chemical equilibrium)  where the joint distribution has converged to a fixed {\emph{stationary}} 
distribution that is no longer changing in time.  A sufficient condition for a stationary distribution to exist is that the reaction rates satisfy constraints known as {\emph{circuit conditions}} \cite{feinberg89,joshi15} or that the network has {\emph{deficiency zero}} \cite{anderson2010, levien2017, angeli2009}.


\subsection{A Two-Species Network}
\slabel{twospecies}
As a simple example, take the two-species CRN given in Figure \ref{fig:Net2a} in which a molecule of {\ce{A}} converts to two molecules of {\ce{B}} and two molecules of {\ce{B}} convert to a molecule of {\ce{A}}. Such a network could serve, for example, as a model of protein dimerization and dissociation with {\ce{A}} representing dimers and {\ce{B}} representing monomers.
\vspace{0.15in}
\tikzstyle{block} = [draw,draw=white,fill=white!20,minimum size=2em]
\begin{figure}[!h]
\vspace{-0.2in}
\begin{center}
\begin{tikzpicture}[auto, node distance=1cm,>=latex']
    \node [block, name=A] (A) { \mbox{\large {\ce{A}}} };
    \node [block, right of=A,node distance =1.75cm] (B) { \mbox{\large $2B$} };
    \draw[thick,->] ([yshift= 0.08cm]A.east) to [out=0,in=180] node[auto] {$k_{1}$} ([yshift= 0.08cm]B.west);
    \draw[thick,->] ([yshift= -0.08cm]B.west) to [out=180,in=0] node[auto] {$k_{2}$} ([yshift= -0.08cm]A.east);       
\end{tikzpicture}
\caption{A Two-Species Network} 
\label{fig:Net2a}
\end{center}
\end{figure}
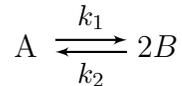

We consider modelling $X(t) = (X_{A}(t),X_{B}(t))$ where $X_{A}(t)$ is the number of molecules of species {\ce{A}} at time $t$ and $X_{B}(t)$ is the number of molecules of species {\ce{B}}. At any instant of time when $X(t)=(x_{A},x_{B})$, a molecule of {\ce{A}} converts to two molecules of {\ce{B}} after an exponential amount of time with rate $k_{1} x_{A}$ and two molecules of {\ce{B}} convert to a molecule of {\ce{A}} after an exponential amount of time with rate  $ k_{2} \binom{x_{B}}{2} = \frac{1}{2} k_{2} x_{B} (x_{B}-1)$.

Note that the network has the conservation law meaning, in this case, that the quantity $2X_{A}(t)+X_{B}(t)$ is constant for all time. We will call this constant $N$. We can now completely describe $X(t)$ by just considering $X_{A}(t)$. $\{X_{A}(t)\}$ is a birth-and-death process on the state space $\{0,1,2,\hdots,n\}$ where $n=N/2$  in the case that $N$ is even and $n=(N-1)/2$ in the case that $N$ is odd. 

The birth and death rates are 
\begin{align}
\lambda_{x_{A}} = \left\{
\begin{array}{lcl}
\frac{1}{2} k_{2} (N-2x_{A})(N-2x_{A}-1) &,& \mbox{if} \,\,\, x_{A}=0,1, \hdots,n-1 \\
0 &,& \mbox{otherwise}
\end{array}
\right.
\label{e:birthrate}
\end{align}
and
\begin{align}
\mu_{x_{A}} = \left\{
\begin{array}{lcl}
k_{1} x_{A} &,& \mbox{if} \,\,\, x_{A}=1,2,\hdots,n\\
0 &,& \mbox{otherwise}
\end{array}
\right.
\label{e:deathrate}
\end{align}

We can then use (\ref{e:bdstat}) to easily find the exact values of $\pi_{0},\pi_{1}, \pi_{2}, \hdots \pi_{n}$ which has the product form
\begin{align}
    \pi_{j} 
        &= \pi_{0} \prod_{x_{A}=1}^{j} \frac{ k_{2} (N-2x_{A}+2)(N-2x_{A}+1) }{ 2 k_{1} x_{A} }
        \label{Net2_prodform}
\end{align}
with the constraint that $\sum_{j=0}^{n} \pi_{j}=1$.

Alternatively, we can compute the values $\pi_{0},\pi_{1}, \pi_{2}, \hdots, \pi_{n}$ using Algorithm \ref{alg:perfect_BD}. As in the exact calculation, the transition rates for the simulation are given by mass action kinetics and are identical to the rates given by  \eqref{e:birthrate} and \eqref{e:deathrate}. 
Note that $\lambda_{x_{A}}$ is decreasing in $x_{A}$ and that $\mu_{x_{A}}$ is increasing. Thus, we have that
$$
\lambda^{*} = \max_{0 \leq x_{A} \leq n} \{\lambda_{x_{A}}\} = \lambda_{0} \,\,\,\, \mbox{and} \,\,\,\, \mu^{*} = \max_{0 \leq x_{A} \leq n} \{\mu_{x_{A}}\} = \mu_{n}.
$$ 
are the transition rates of the maximum rate process $\{Y(t)\}$.
 
\begin{figure}[!htb]
    \centering
    \includegraphics[width = 0.4\textwidth]{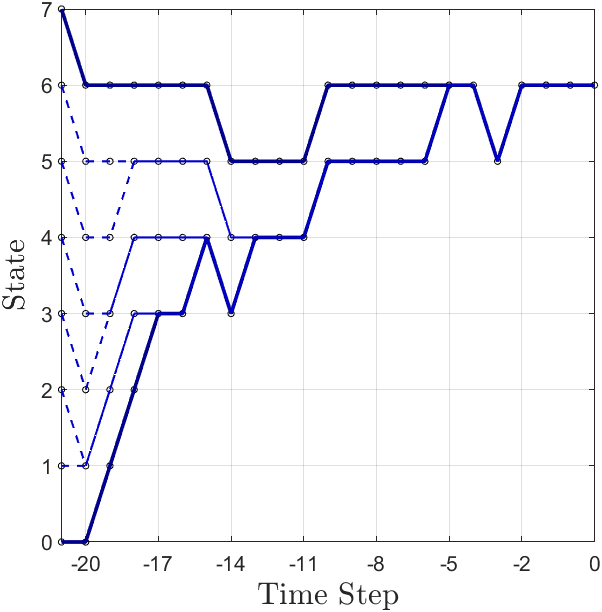}
    \caption{Coupling of monotone process: stochastic dominance reduces the number of paths that must be constructed for two-species network. The result is a single draw from the stationary distribution $X_{A} \sim \pi_{A}$}
    \label{fig:Net2_StochMon}
\end{figure}

 As illustrated in Figure \ref{fig:Net2_StochMon}, 
simulation of the distribution may be done using only the paths from the ``top" ($X_{A} = n$) and ``bottom" ($X_{A} = 0$) of the space since intermediate paths will remain sandwiched between these bounding chains. Indeed, consider two sample paths currently at points $x_{A}$ and $x_{A}^{\prime}$ with $x_{A} < x_{A}^{\prime}$. Suppose that the maximum rate process causes a birth in the system. Each of these two paths then has the potential for a birth. Due to the fact that $\lambda_{x_{A}}$ is decreasing, the probability for a birth for the path at  $x_{A}$ is larger than the probability for a birth for the path at $x_{A}^{\prime}$ and the paths remain ordered. Similarly, suppose that the maximum rate processes causes a death in the system. Each of these two paths has the potential for a death but because $\mu_{x_{A}}$ is increasing the probability for a death for the path at $x_{A}^{\prime}$ is larger than the probability for a death for the path at $x_{A}$ and the paths still remain ordered.

 
\subsubsection{Results} 
To demonstrate the Perfect-Birth-and-Death algorithm, we simulated $50,000$ values from the stationary distribution $\pi = (\pi_{0}, \pi_{1}, \hdots, \pi_{n})$ for the simple two-species network that results when $n=7$ and $N=15$. In this particular example, we follow the ``top" and ``bottom" paths started from $X_{A} = 7$ and $X_{A}=0$, respectively. The results are compared to the exact solution computed from \eqref{e:bdstat} and are summarized in Table \ref{fig:a2bresults}.

\begin{figure}[!h]
    \centering
    \begin{minipage}{.31\textwidth}
        \centering
        \includegraphics[width=\linewidth, height=0.25\textheight]{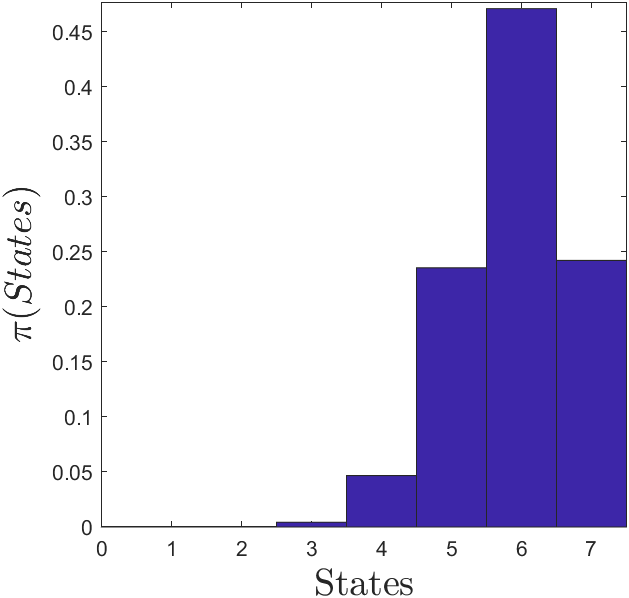} 
    \end{minipage}%
    \hfill
    \begin{minipage}{0.31\textwidth}
        \centering
        \includegraphics[width=\linewidth, height=0.25\textheight]{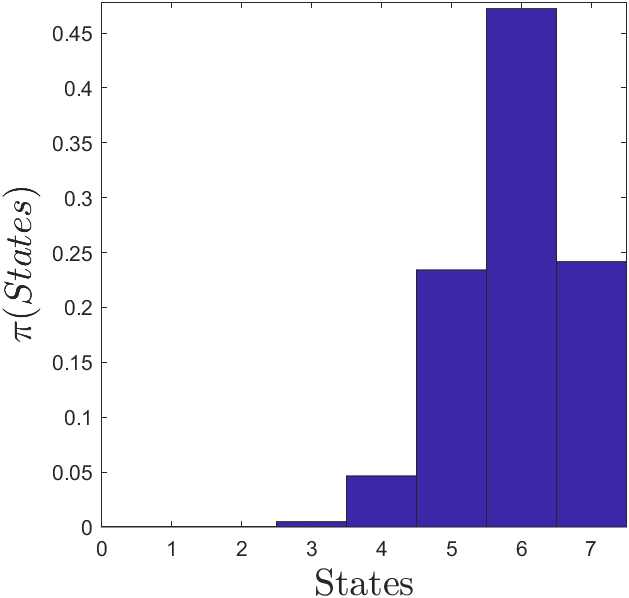} 
    \end{minipage}%
    \hfill
    \begin{minipage}{0.32\textwidth}
        \centering
        \includegraphics[width=\linewidth, height=0.25\textheight]{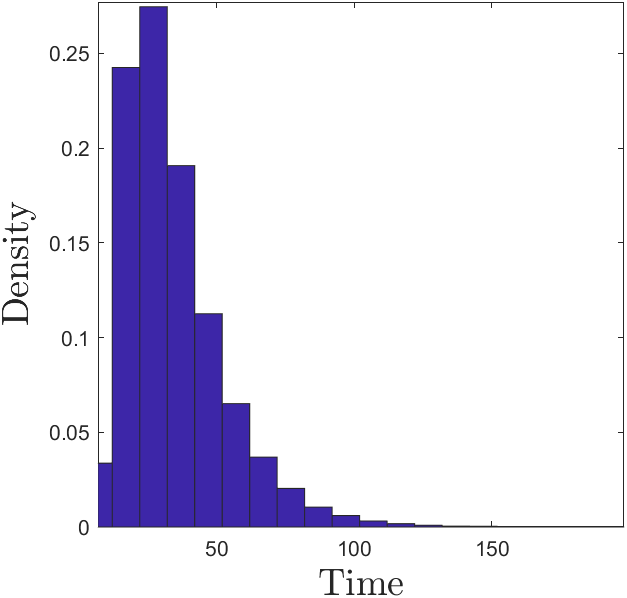} 
    \end{minipage}%
     \caption{Stationary distribution for the two-species network with $k_{1} = 1$, $k_{2} = 1.2$, $n=7$ and $N=15$. \emph{\bf Left} Exact solution computed using \eqref{e:bdstat}. \emph{\bf Center} Perfect simulation results using Algorithm \ref{alg:perfect_BD}. \emph{\bf Right} Histogram of coupling times.}  
    \label{fig:a2bsoln}
\end{figure}

\begin{table}[htb!]
    \centering 
        \begin{tabular}{c||c|c}
           State   &   Exact                    & Simulation \\ \hline
         $X_{A} = 0$ & $ 3.3351 \times 10^{-8} $  &  0       \\
         $X_{A} = 1$ & $ 4.2022 \times 10^{-6} $  &  0  \\
         $X_{A} = 2$ & $ 1.9666 \times 10^{-4} $  &  $1.600 \times 10^{-4} $  \\
         $X_{A} = 3$ & $ 4.3266 \times 10^{-3} $  &  $4.2500 \times 10^{-3} $  \\
         $X_{A} = 4$ & $ 0.04673$                 &  $0.0473$ \\
         $X_{A} = 5$ & $ 0.2355$                  &  $0.2354$  \\
         $X_{A} = 6$ & $ 0.4710$                  &  $0.4704$  \\
         $X_{A} = 7$ & $ 0.2422$                  &  $0.2425$   
        \end{tabular}
    \caption{Stationary probabilities computed for the two-species network with $k_{1} = 1$ and $k_{2} = 1.2$, $n=7$, and $N=15$.}        
    \label{fig:a2bresults}
\end{table}

We computed the \emph{total variation norm distance} between the true and resulting estimated values for $\pi$. In the case of a discrete state space like we have here, this is given by 
\begin{align}
    ||\widehat{\pi}-\pi||_{_{TVN}}
        &= \frac{1}{2} \sum_{j} |\widehat{\pi}_{j}-\pi_{j}|
        \label{e:normerr}
\end{align}
where the sum is taken over all states $j$.

The total variation norm distance between the simulated distribution and the exact probability distribution in this example is $||\widehat{\pi}-\pi||_{_{TVN}} = 0.001637$, which is approximately the 16th percentile of the total variation norm distances for exact samples of size $50,000$ from $\pi$.

\section{Perfect Simulation for General CRNs}
\label{s:perfectchemnet}

Unlike the two-species problem introduced in \Section{networks}, a typical CRN is not adequately described by a birth-and-death process alone. However, our algorithm will extend in a natural way.

To define the transitions, we replace the notion of species ``births" or ``deaths" with chemical reactions $R_{i}$, $i = 1, 2, \hdots, m$. While a simple birth-and-death process involves single step transitions of one species, the $R_{i}$ may change the molecular counts of multiple species by any amount given in the reaction. 

We also generalize the idea of ``birth rates" and ``death rates" simply to ``reaction rates" $\lambda_{i}(x)$, for $i=1,2, \ldots, m$, which are given by mass action kinetics and depend on the current state $x$. We assume that reaction $R_{i}$ has a corresponding reaction rate $\lambda_{i}(x)$ which is defined using a rate constant $k_{i}$. (This is in contrast to standard notation in  the CRN literature where, for example, $k_{i}$ and $k_{-i}$ denote rate constants for a reaction and its reverse reaction, respectively.) Further, we assume that reaction $R_{i}$ is a vector in the sense that a network in state $x$ that undergoes reaction $R_{i}$ moves to state $x+R_{i}$. 

As before, we run paths by thinning a maximum rate process. The maximum rate for reaction $i$ over all possible states of the network is
\begin{align*}
    \lambda_{i}^{*} &= \max_{ x \in \mathbb{S} } \{ \lambda_{i}(x) \}.
\end{align*}
where ${\mathbb{S}}$ is the state space of all vectors of possible combinations of molecular counts.

Let $R= \sum_{i=1}^{m} \lambda_{i}^{*}$ be the rate of the next reaction in the maximum rate process. We now describe our perfect simulation algorithm under the assumption that one can consider sample paths from all possible states of the system. As this is clearly not feasible for most problems, we loosen this restriction in Section \ref{s:michaelis}. As before, we first determine a move for the maximum rate process and then allow the process of interest to either accept or reject this move. The possible moves are any of the reactions $R_{i}, \ i = 1, \hdots, m$.

\begin{algorithm}[H]
\caption{Perfect CRN}
\begin{enumerate}
\item[0.] Choose a fixed integer $N \geq 1$. Set $n=N$.

\item Generate and store $2N$ independent  and identically distributed random variables $U_{-n}, U_{-n+1}, \ldots, U_{-n+N-1}$ and $V_{-n}, V_{-n+1}, \ldots, V_{-n+N-1}$ that are uniformly distributed over the interval $(0,1)$.
    
\item Start sample paths at time $-n$ by setting
$$
X_{i}^{(-n)} = i
$$
for all $i \in S$.

\item For each time step $ \, t = n, n-1, \hdots, 1$, determine the type of transition for the maximum rate process and consider it as a possible transition for all sample paths as follows.
    
    \begin{itemize}
    \item[] If $U_{-t} \leq \min \{k: \sum_{j=i}^{k} \lambda_{j}^{*} /R\}$ then there is a type $k$ transition in the maximum rate process. In this case, all paths have the potential to undergo the reaction $R_{k}$.

    Specifically, for $i = 1, 2, \hdots, m$ let $x = X_{i}^{(-t)}$ and assign
    $$
        X_{i}^{(-t+1)} = x + R_{k}
    $$
    whenever $V_{-t} \leq \lambda_{k}(x)/\lambda_{k}^{*}$. Otherwise, $X_{i}^{(-t+1)} = x$.
    \end{itemize}

\item If the $X_{i}^{(0)}$ are equal for all $i \in S$, stop the algorithm. This common value is a perfect draw from the stationary distribution for the birth-and-death chain.

Otherwise, set n = n+N. and return to Step 1.
\end{enumerate}
\label{alg:perfect_crn}
\end{algorithm}

Again, while we are moving back in time-blocks of length $N$, in Step 2 we are always going all the way forward to time $0$ and are reusing previously generated blocks of $U$'s and $V$'s.

\subsection{The Reversible Michaelis-Menten Model}
\slabel{michaelis}

We now return to the Reversible Michaelis-Menten (RMM) model first introduced in \Section{int} and depicted in Figure \ref{fig:RMM}.

\tikzstyle{block} = [draw,draw=white,fill=white!20,minimum size=2em]
\begin{figure}[h]
\begin{center}
\begin{tikzpicture}[auto, node distance=1cm,>=latex']
    \node [block, name=C] (C) {\mbox{ \large $C$} };
    \node [block, left of=C, node distance=2cm] (SE) {\mbox{ \large$S+E$} };
    \node [block, right of=C,node distance =2cm] (PE) {\mbox{ \large$P+E$} };
    
    \draw[thick,->] ([yshift= 0.08cm]SE.east) to [out=0,in=180] node[auto] {$k_{1}$} ([yshift= 0.08cm]C.west);
    \draw[thick,->] ([yshift= -0.08cm]C.west) to [out=180,in=0] node[auto] {$k_{2}$} ([yshift= -0.08cm]SE.east);
    \draw[thick,->] ([yshift= 0.08cm]C.east) to [out=0,in=180] node[auto] {$k_{3}$} ([yshift= 0.08cm]PE.west);
    \draw[thick,->] ([yshift= -0.08cm]PE.west) to [out=180,in=0] node[auto] {$k_{4}$} ([yshift= -0.08cm]C.east);        
\end{tikzpicture}
\caption{The Reversible Michaelis-Menten Network}
\label{fig:RMM}
\end{center}
\end{figure}
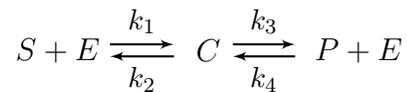

Following the notation in the previous example, we wish to model $X(t) = (X_{C}(t), X_{S}(t), X_{E}(t), X_{P}(t))$ where the $X(t)$ are the numbers of molecules of species {\ce{C}}, {\ce{S}}, {\ce{E}}, and {\ce{P}} for any time $t \geq 0$. Notice we can model $X(t)$ by just modelling e.g. $(X_{C}(t), X_{P}(t))$ since $N := X_{C}(t) + X_{S}(t) + X_{P}(t)$ and $M:= X_{C}(t) + X_{E}(t)$ must be constant for all time.

In contrast to the two species model, the state space of the RMM network is at least two-dimensional and there is no longer a clear notion of ``top" or ``bottom" paths that can be used in simulation to sandwich intermediate paths. 
However, stochastic monotonicity in an ordered state space can be generalized to a \emph{subset} of the state space such that coupling of paths constructed for states in the subset implies coupling of all paths even if they do not remain sandwiched along the way. To efficiently compute perfect draws from the stationary distribution of $\{X_{C},X_{P}\}$, we must find this subset which we will denote $\mathbb{S}_{C}$.


To motivate our choice of $\mathbb{S}_{C}$, recall that ordering of the reaction rates in the two-species network resulted in chains that were also ordered. Thus, coupling of paths that started from $X_{A} = 0$ and $X_{A} = n$ implied coupling of all paths because these states maximized the birth and death rates, respectively, throughout the simulation.

Extending this approach to the RMM problem, we found a subset $\mathbb{S}_{C} \subset \mathbb{S}$ that maximizes the transition rates $\lambda_{i}(x)$ where $x \in \mathbb{S}$ and has components $x = (x_{C}, x_{P})$. These rates are given by mass action kinetics as before
\begin{align*}
    \lambda_{1}(x_{C}, x_{P}) &= k_{1} (N - x_{C} - x_{P})(M - x_{C}) \\ 
    \lambda_{2}(x_{C}, x_{P}) &= k_{2} \, x_{C}  \\
    \lambda_{3}(x_{C}, x_{P}) &= k_{3} \, x_{C} \\
    \lambda_{4}(x_{C}, x_{P}) &= k_{4} \, x_{P} (M - x_{C}) 
\end{align*} 

\begin{figure}[!t]
    \centering
    \includegraphics[width = 0.35\textwidth]{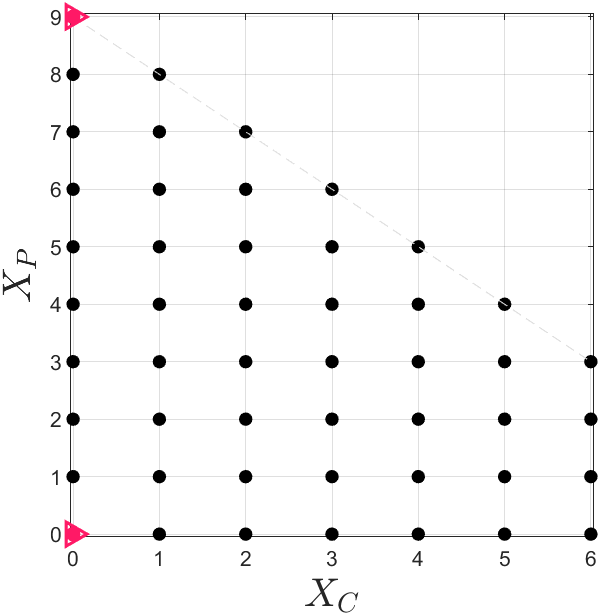}
    \caption{The state space $\mathbb{S}$ when $N=9$ and $M=6$. The states in $\mathbb{S}_{C}$ are marked with solid pink triangles and are in the ``corners" of $\mathbb{S}$.}    
    \label{fig:RMM_state}
\end{figure}

and are maximized by the candidate subset of states
$$
    \big\{ (0,0), \, (0,N), \, (M,x_{p}) \big\}, \qquad 0 \leq x_{p} \leq N-M
$$

Although one could show that paths starting from this subset maximize the reaction rates throughout the simulation and result in coupling of all paths, we are able to say something stronger. Choose the subset to be the collection of states $\ds \mathbb{S}_{C} = \big \{ (0,0), \ (0,N) \big\} $ depicted in Figure \ref{fig:RMM_state}. Under any arbitrary sequence of reaction steps, coupling of sample paths for this subset guarantees coupling of all paths for $\mathbb{S}$. (A proof of this statement is provided in Appendix A.)



\subsubsection{Results}
\begin{figure}[!h]
    \centering
    \begin{minipage}{.3\textwidth}
        \centering
        \includegraphics[width=\linewidth, height=0.25\textheight]{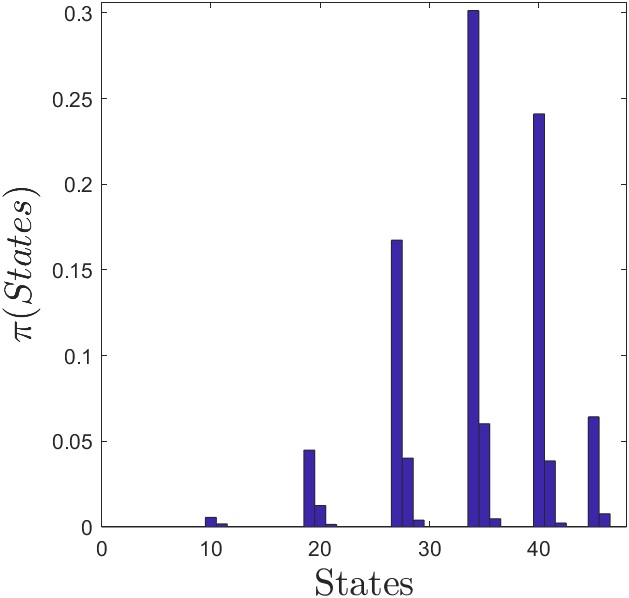} 
    \end{minipage}%
    \hfill
    \begin{minipage}{0.3\textwidth}
        \centering
        \includegraphics[width=\linewidth, height=0.25\textheight]{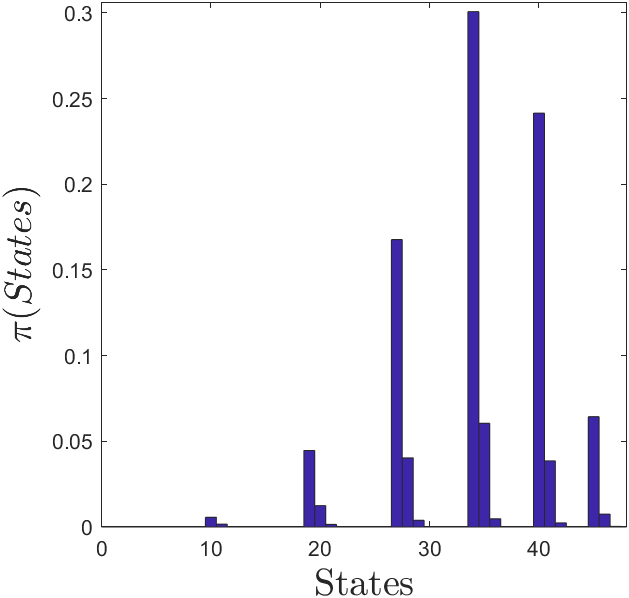} 
    \end{minipage}%
    \hfill
    \begin{minipage}{0.3\textwidth}
        \centering
        \includegraphics[width=\linewidth, height=0.25\textheight]{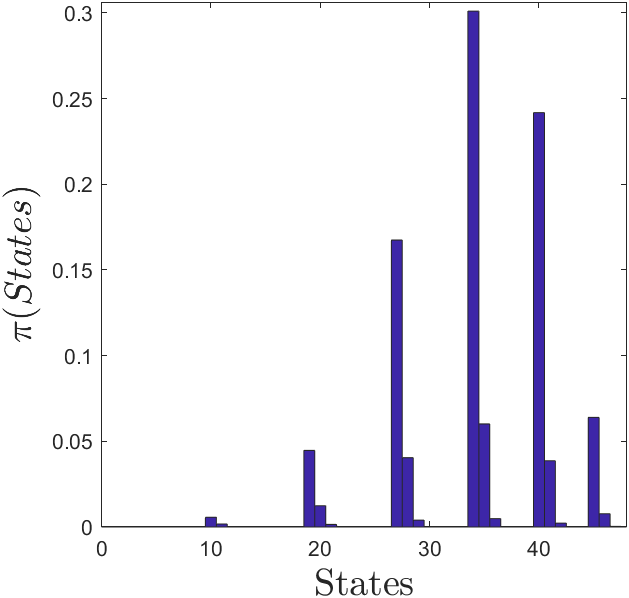} 
    \end{minipage}%
     \caption{Stationary distribution for the RMM network with $k_{1} = 10$, $k_{2} = 1$, $k_{3} = 5$, $k_{4} = 2$, $N=9$, and $M=3$. \emph{\bf Left} Exact solution computed using the CME equation. \emph{\bf Center} Perfectly simulated solution using all states in $\mathbb{S}$. \emph{\bf Right} Perfectly simulated solution using only states in $\mathbb{S}_{C}$.}  
    \label{fig:RMMsoln1}
\end{figure}

To illustrate the Perfect-CRN algorithm, we simulated the stationary probability distribution $\pi = (\pi_{0}, \pi_{1}, \hdots, \pi_{n})$ for the RMM network with $N=9$ and $M=6$. In this particular example, we compare two different simulations to the exact solution computed with \eqref{e:CME}. The first simulation started paths from the ``corner states" in $\mathbb{S}_{C}$, $\{ (0,0), \ (0,9)\}$, while the second simulation started paths from every state in $\mathbb{S}$ but used the same random number streams $\{U_{-n}\}$ and $\{V_{-n}\}$. Both simulations were terminated after 50,000 samples were drawn from the stationary distribution of $(X_{C},X_{P})$. The results of the two simulations are compared to the exact solution in Figure \ref{fig:RMMsoln1}, and it is clear that the simulated results agree with the exact probabilities.


The total variation norm distance between the simulated distribution and the exact probability distribution is $||\widehat{\pi}-\pi||_{_{TVN}} = 0.001276$ for this simulation. As before, this is well within a range of values to be expected for a correct simulation of $50,000$ values.

\begin{figure}[!t]
    \centering
    \begin{minipage}{.35\textwidth}
        \flushright
        \includegraphics[width=0.95\linewidth, height=0.25\textheight]{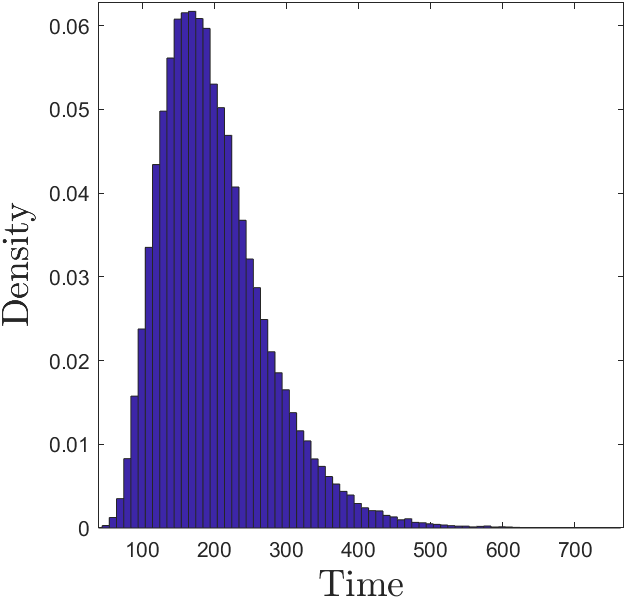} 
    \end{minipage}
    \hfill
    \begin{minipage}{0.35\textwidth}
        \flushleft
        \includegraphics[width=0.95\linewidth, height=0.25\textheight]{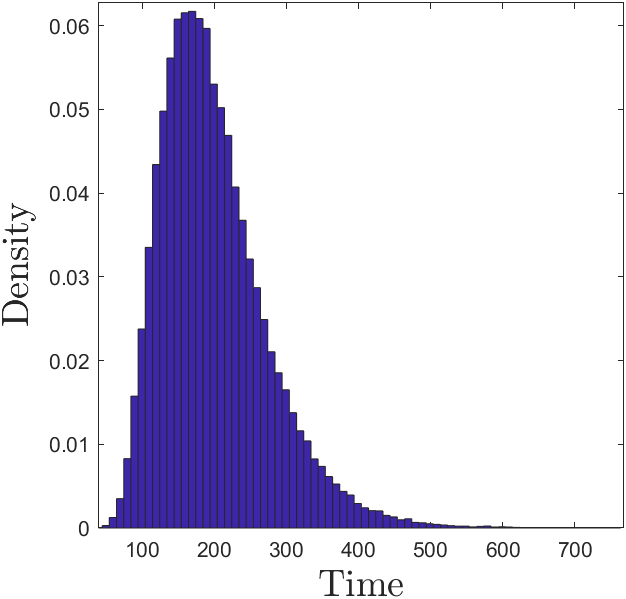} 
    \end{minipage}
     \caption{Distribution of backward coupling times for the RMM network with reaction rate constants $k_{1} = 10$, $k_{2} = 1$, $k_{3} = 5$, and $k_{4} = 2$, $N=9$, and $M=6$. \emph{\bf Left} Backward coupling times when paths are constructed for $X \in \mathbb{S}$. \emph{\bf Right} Backward coupling times when paths are constructed for $X \in \mathbb{S}_{C}$.}    
    \label{fig:RMMtimes}
\end{figure}

In addition to comparing the simulated probabilities, we compared the distributions of backward coupling times $T$ for the two simulations. When the same random number streams $ \{ U_{-n} \}$ and $\{ V_{-n} \}$ are used in the Perfect-CRN algorithm for both simulations, the distribution of coupling times and their statistics are indistinguishable, as they should be. These results are summarized in Table \ref{table:RMMcouptimes} and Figure \ref{fig:RMMtimes}.

\vspace{0.25in}
\begin{table}[!h]
    \centering 
        \begin{tabular}{c||c|c|c}
           Method        & Mean     & Median & Mode \\ \hline 
           $X \in \mathbb{S}$     & 200.1895 & 188    & 176  \\ \hline 
           $X \in \mathbb{S}_{C}$ & 200.1895 & 188    & 176
        \end{tabular}
    \caption{Comparing the coupling times of the perfect simulations of the RMM network.}        
    \label{table:RMMcouptimes}
\end{table}

\subsection{Larger Linear Models}
\label{s:BigLins}
So far we have considered CRNs with coupled, linear, first- and second-order reversible reactions. The examples provided so far have only included networks of two or four reactions. 
While our backward coupling algorithm is applicable to more general networks, simulations suggest that the ability for us to follow corner points only require  that the network be \emph{linear} as well as \emph{reversible}. However, the restriction to first- and second-order reactions can be relaxed if every chemical species in the network is distinct\footnote{The restriction to unique chemical species ensures that the state space $\mathbb{S}$ is closed for an arbitrary  $X(0)$.}. The final examples consider four such networks which are given in Figure \ref{fig:BigLins}.


\begin{figure}[htb!]
\begin{minipage}{\textwidth}
\centering
\begin{minipage}{0.75\textwidth}
\begin{flushright}
\tikzstyle{block} = [draw,draw=white,fill=white!20,minimum size=2em]
\begin{tikzpicture}[auto, node distance=1cm,>=latex']
    \node [block, name=AB] (AB) {\mbox{ \large $A+B$} };
    \node [block, right of=AB, node distance=2.25cm] (CD) {\mbox{ \large$C+D$} };
    \node [block, right of=CD,node distance =2.25cm] (EF) {\mbox{ \large$E+F$} };
    \node [block, right of=EF,node distance =2.25cm] (GH) {\mbox{ \large$G+H$} };
    \draw[thick,->] ([yshift= 0.08cm]AB.east) to [out=0,in=180] node[auto] {$k_{1}$} ([yshift= 0.08cm]CD.west);
    \draw[thick,->] ([yshift= -0.08cm]CD.west) to [out=180,in=0] node[auto] {$k_{2}$} ([yshift= -0.08cm]AB.east);
    \draw[thick,->] ([yshift= 0.08cm]CD.east) to [out=0,in=180] node[auto] {$k_{3}$} ([yshift= 0.08cm]EF.west);
    \draw[thick,->] ([yshift= -0.08cm]EF.west) to [out=180,in=0] node[auto] {$k_{4}$} ([yshift= -0.08cm]CD.east);     
    \draw[thick,->] ([yshift= 0.08cm]EF.east) to [out=0,in=180] node[auto] {$k_{5}$} ([yshift= 0.08cm]GH.west);
    \draw[thick,->] ([yshift= -0.08cm]GH.west) to [out=180,in=0] node[auto] {$k_{6}$} ([yshift= -0.08cm]EF.east); 
\end{tikzpicture}
\end{flushright}
\end{minipage}%
\hfill
\begin{minipage}{0.15\textwidth}
    \begin{flushright} 
        $\ds (a)$
    \end{flushright}
\end{minipage}
\end{minipage} %

\vfill

\begin{minipage}{\textwidth}
\centering
\begin{minipage}{0.75\textwidth}
\begin{flushright}
\tikzstyle{block} = [draw,draw=white,fill=white!20,minimum size=2em]
\begin{tikzpicture}[auto, node distance=2cm,>=latex']
    \node [block, name=A] (A) {\mbox{ \large $2A$} };
    \node [block, right of=A, node distance=2cm] (BC) {\mbox{ \large$B+2C$} };
    \node [block, right of=BC, node distance =2.4cm] (DE) {\mbox{ \large$D+4E$} };
    \node [block, right of=DE, node distance =2.4cm] (FG) {\mbox{ \large$2F+G$} };
    \draw[thick,->] ([yshift= 0.08cm]A.east) to [out=0,in=180] node[auto] {$k_{1}$} ([yshift= 0.08cm]BC.west);
    \draw[thick,->] ([yshift= -0.08cm]BC.west) to [out=180,in=0] node[auto] {$k_{2}$} ([yshift= -0.08cm]A.east);
    \draw[thick,->] ([yshift= 0.08cm]BC.east) to [out=0,in=180] node[auto] {$k_{3}$} ([yshift= 0.08cm]DE.west);
    \draw[thick,->] ([yshift= -0.08cm]DE.west) to [out=180,in=0] node[auto] {$k_{4}$} ([yshift= -0.08cm]BC.east);     
    \draw[thick,->] ([yshift= 0.08cm]DE.east) to [out=0,in=180] node[auto] {$k_{5}$} ([yshift= 0.08cm]FG.west);
    \draw[thick,->] ([yshift= -0.08cm]FG.west) to [out=180,in=0] node[auto] {$k_{6}$} ([yshift= -0.08cm]DE.east); 
\end{tikzpicture}
\end{flushright}
\end{minipage}%
\hfill
\begin{minipage}{0.15\textwidth}
    \begin{flushright} 
        $\ds (b)$
    \end{flushright}
\end{minipage}
\end{minipage}

\vfill

\begin{minipage}{\textwidth}
\centering
\begin{minipage}{0.9\textwidth}
\begin{flushright}
\tikzstyle{block} = [draw,draw=white,fill=white!20,minimum size=2em]
\begin{tikzpicture}[auto, node distance=1.75cm,>=latex']
    \node [block, name=A] (A) {\mbox{ \large $2A$} };
    \node [block, right of=A, node distance=2cm] (BC) {\mbox{ \large$B+C$} };
    \node [block, right of=BC, node distance =2.85cm] (DEF) {\mbox{ \large$D+E+F$} };
    \node [block, right of=DEF, node distance =2.5cm] (G) {\mbox{ \large$G$} };
    \node [block, right of=G, node distance =2.1cm] (HI) {\mbox{ \large$2H+I$} };
    \node [block, right of=HI, node distance =2.4cm] (JK) {\mbox{ \large$J+K$} };
    \draw[thick,->] ([yshift= 0.08cm]A.east) to [out=0,in=180] node[auto] {$k_{1}$} ([yshift= 0.08cm]BC.west);
    \draw[thick,->] ([yshift= -0.08cm]BC.west) to [out=180,in=0] node[auto] {$k_{2}$} ([yshift= -0.08cm]A.east);
    \draw[thick,->] ([yshift= 0.08cm]BC.east) to [out=0,in=180] node[auto] {$k_{3}$} ([yshift= 0.08cm]DEF.west);
    \draw[thick,->] ([yshift= -0.08cm]DEF.west) to [out=180,in=0] node[auto] {$k_{4}$} ([yshift= -0.08cm]BC.east);     
    \draw[thick,->] ([yshift= 0.08cm]DEF.east) to [out=0,in=180] node[auto] {$k_{5}$} ([yshift= 0.08cm]G.west);
    \draw[thick,->] ([yshift= -0.08cm]G.west) to [out=180,in=0] node[auto] {$k_{6}$} ([yshift= -0.08cm]DEF.east); 
    \draw[thick,->] ([yshift= 0.08cm]G.east) to [out=0,in=180] node[auto] {$k_{7}$} ([yshift= 0.08cm]HI.west);
    \draw[thick,->] ([yshift= -0.08cm]HI.west) to [out=180,in=0] node[auto] {$k_{8}$} ([yshift= -0.08cm]G.east);    
    \draw[thick,->] ([yshift= 0.08cm]HI.east) to [out=0,in=180] node[auto] {$k_{9}$} ([yshift= 0.08cm]JK.west);
   \draw[thick,->] ([yshift= -0.08cm]JK.west) to [out=180,in=0] node[auto] {$k_{10}$} ([yshift= -0.08cm]HI.east);        
\end{tikzpicture}
\end{flushright}
\end{minipage}%
\hfill
\begin{minipage}{0.1\textwidth}
    \begin{flushright} 
        $\ds (c)$
    \end{flushright}
\end{minipage}
\end{minipage}

\vfill

\begin{minipage}{\textwidth}
\centering
\begin{minipage}{0.9\textwidth}
\begin{flushright}
\tikzstyle{block} = [draw,draw=white,fill=white!20,minimum size=2em]
\begin{tikzpicture}[auto, node distance=1.75cm,>=latex']
    \node [block, name=A] (A) {\mbox{ \large $2A$} };
    \node [block, right of=A, node distance=2.1cm] (BC) {\mbox{ \large$B+C$} };
    \node [block, right of=BC, node distance =2.5cm] (DE) {\mbox{ \large$D+E$} };
    \node [block, right of=DE, node distance =2.25cm] (F) {\mbox{ \large$F$} };
    \node [block, right of=F, node distance =2cm] (G) {\mbox{ \large$2G$} };
    \node [block, right of=G, node distance =2cm] (H) {\mbox{ \large$H$} };
    \node [block, right of=H, node distance =2.1cm] (IJ) {\mbox{ \large$I+J$} };
    \draw[thick,->] ([yshift= 0.08cm]A.east) to [out=0,in=180] node[auto] {$k_{1}$} ([yshift= 0.08cm]BC.west);
    \draw[thick,->] ([yshift= -0.08cm]BC.west) to [out=180,in=0] node[auto] {$k_{2}$} ([yshift= -0.08cm]A.east);
    \draw[thick,->] ([yshift= 0.08cm]BC.east) to [out=0,in=180] node[auto] {$k_{3}$} ([yshift= 0.08cm]DE.west);
    \draw[thick,->] ([yshift= -0.08cm]DE.west) to [out=180,in=0] node[auto] {$k_{4}$} ([yshift= -0.08cm]BC.east);     
    \draw[thick,->] ([yshift= 0.08cm]DE.east) to [out=0,in=180] node[auto] {$k_{5}$} ([yshift= 0.08cm]F.west);
    \draw[thick,->] ([yshift= -0.08cm]F.west) to [out=180,in=0] node[auto] {$k_{6}$} ([yshift= -0.08cm]DE.east); 
    \draw[thick,->] ([yshift= 0.08cm]F.east) to [out=0,in=180] node[auto] {$k_{7}$} ([yshift= 0.08cm]G.west);
    \draw[thick,->] ([yshift= -0.08cm]G.west) to [out=180,in=0] node[auto] {$k_{8}$} ([yshift= -0.08cm]F.east);    
    \draw[thick,->] ([yshift= 0.08cm]G.east) to [out=0,in=180] node[auto] {$k_{9}$} ([yshift= 0.08cm]H.west);
   \draw[thick,->] ([yshift= -0.08cm]H.west) to [out=180,in=0] node[auto] {$k_{10}$} ([yshift= -0.08cm]G.east);     
    \draw[thick,->] ([yshift= 0.08cm]H.east) to [out=0,in=180] node[auto] {$k_{9}$} ([yshift= 0.08cm]IJ.west);
   \draw[thick,->] ([yshift= -0.08cm]IJ.west) to [out=180,in=0] node[auto] {$k_{10}$} ([yshift= -0.08cm]H.east);        
\end{tikzpicture}
\end{flushright}
\end{minipage}%
\hfill
\begin{minipage}{0.1\textwidth}
    \begin{flushright} 
        $\ds (d)$
    \end{flushright}
\end{minipage}
\end{minipage}
\caption{Larger Linear Networks}
\label{fig:BigLins}
\end{figure}  

As before, we model the joint probability distribution $X(t) = (X_{A}(t), X_{B}(t), X_{C}(t), \hdots)$ of each network, where the $X_{i}(t)$ is the number of molecules of species $i = A, B, C, \hdots$ for any time $t \geq 0$, and draw from the stationary distribution $\pi(X)$ using the Perfect-CRN algorithm.

For these larger networks, it is useful to find a subset of states whose paths will bound all paths in the state space. Unlike the previous examples, in which the bounding states $\mathbb{S}_{C}$ geometrically were either the ``top" and ``bottom" states of a one-dimensional state space or the ``corner" states of a two-dimensional state space, the bounding states for the CRNs in Figure \ref{fig:BigLins}(a)-(d) do not have a clear geometric meaning. Following the intuition from the previous examples, we simply define the subset of bounding states for these models to be those states which maximize the transition rates $\lambda_{i}(x)$. For consistency, we will continue to refer to these bounding states as ``corner" states, the collection of  which we denote by $\mathbb{S}_{C}$.

\subsubsection{Results for the Large Linear Networks \ref{fig:BigLins}(a)-(d)}
For each of the CRNs (a), (b), and (c) from Figure \ref{fig:BigLins}, we simulated $50,000$ draws from the stationary distribution $\pi$ using the Perfect-CRN algorithm. For the network depicted in (d), we simulated $150,000$ values as there were more probabilities to be estimated. As before, we drew these samples by constructing paths starting only from the corner states $\mathbb{S}_{C}$. Reusing the random number streams $\{U_{-n}\}$ and $\{V_{-n}\}$ from this first simulation, we then also drew samples by constructing paths starting from every state in $\mathbb{S}$. 

\begin{figure}[!h]
    \centering
    \begin{minipage}{0.25\textwidth}
        \centering
        \includegraphics[width=\linewidth, height=0.2\textheight]{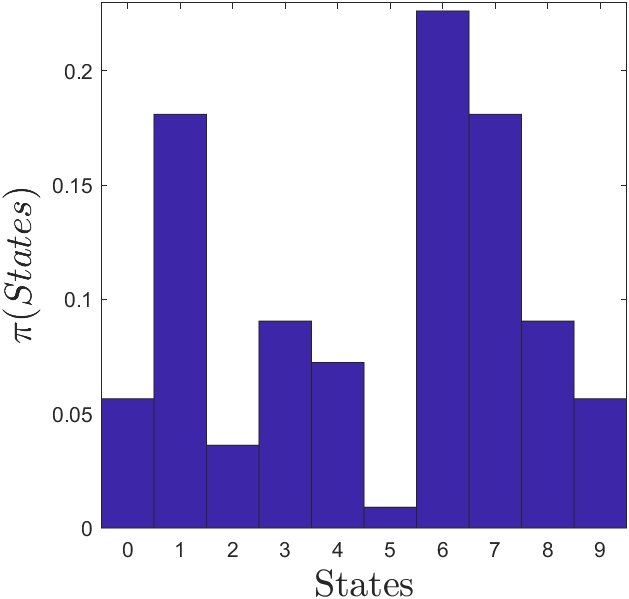} 
    \end{minipage}%
    \hfill
    \begin{minipage}{0.25\textwidth}
        \centering
        \includegraphics[width=\linewidth, height=0.2\textheight]{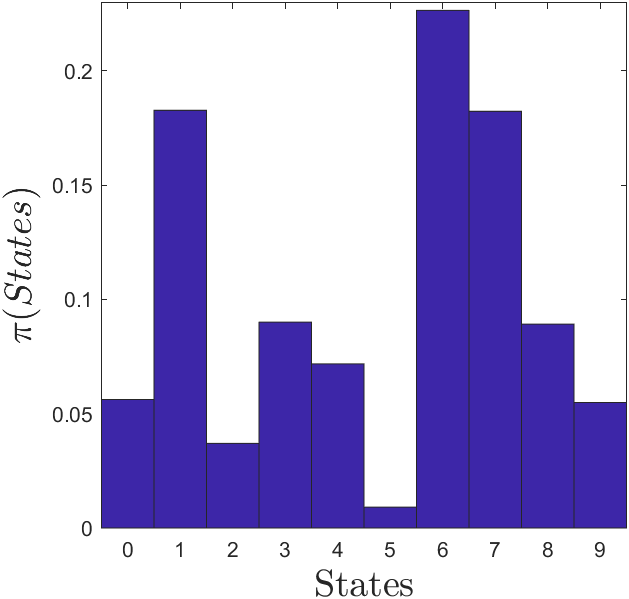} 
    \end{minipage}%
    \hfill
    \begin{minipage}{0.25\textwidth}
        \centering
        \includegraphics[width=\linewidth, height=0.2\textheight]{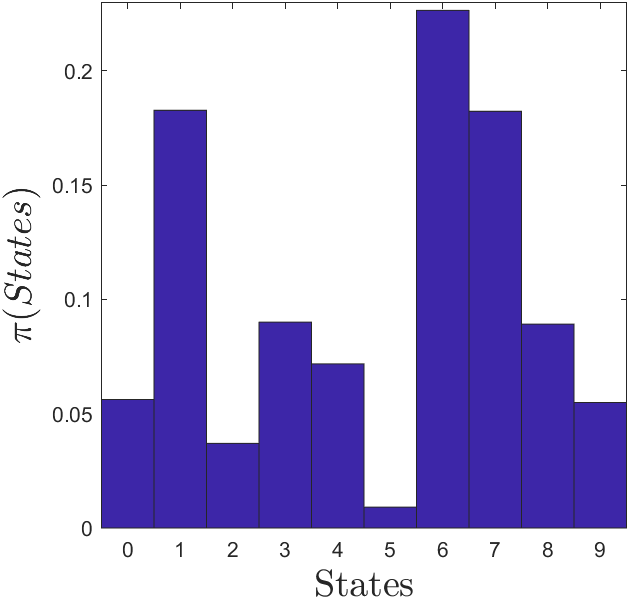} 
    \end{minipage}%
     \caption{Stationary distribution for the Network \ref{fig:BigLins}(a) with $k_{1} = 1$, $k_{2} = 2.5$, $k_{3} = 2$, $k_{4} = 1$, $k_{5}=1.75$, and $k_{6}=1.4$ and a state space that can be generated around the point where there are $2$ molecules of $A$ and $B$, and the remaining molecular counts equal to zero. The total variation norm distance between $\widehat{\pi}$ and $\pi$ was $0.004276$. \emph{\bf Left} Exact solution computed using \eqref{e:CME}. \emph{\bf Center} Perfectly simulated solution using $X \in \mathbb{S}$. \emph{\bf Right} Perfectly simulated solution using only $X \in \mathbb{S}_{C}$.}    

\vspace{0.15in}
    \begin{minipage}{.25\textwidth}
        \centering
        \includegraphics[width=\linewidth, height=0.2\textheight]{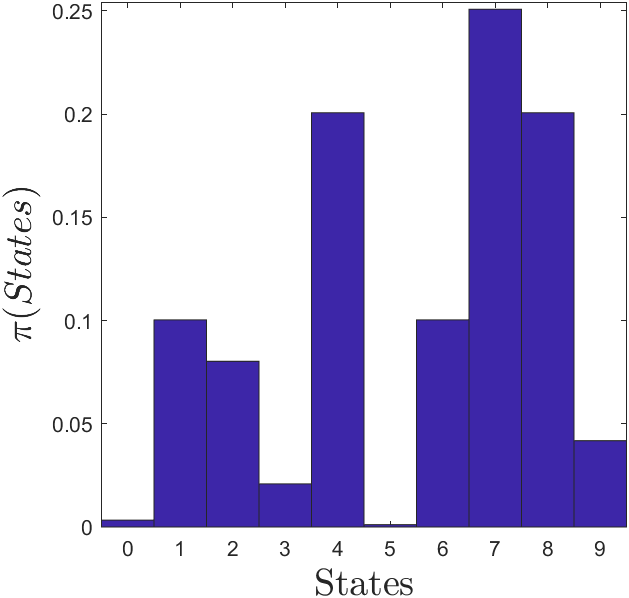} 
    \end{minipage}%
    \hfill
    \begin{minipage}{0.25\textwidth}
        \centering
        \includegraphics[width=\linewidth, height=0.2\textheight]{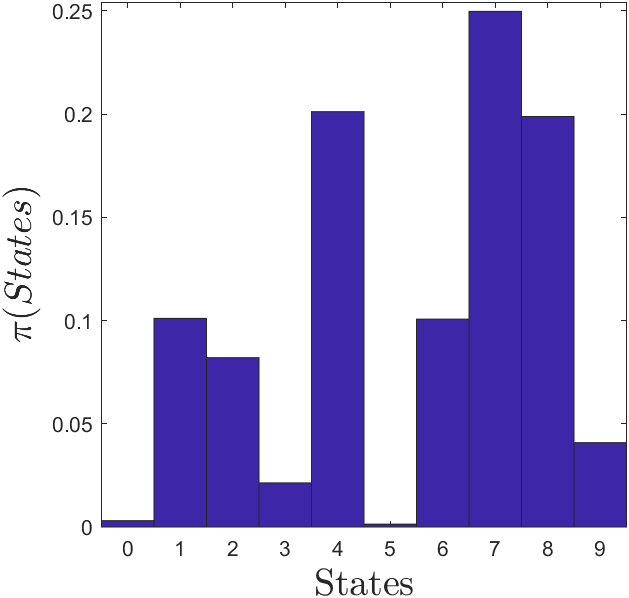} 
    \end{minipage}%
    \hfill
    \begin{minipage}{0.25\textwidth}and
        \centering
        \includegraphics[width=\linewidth, height=0.2\textheight]{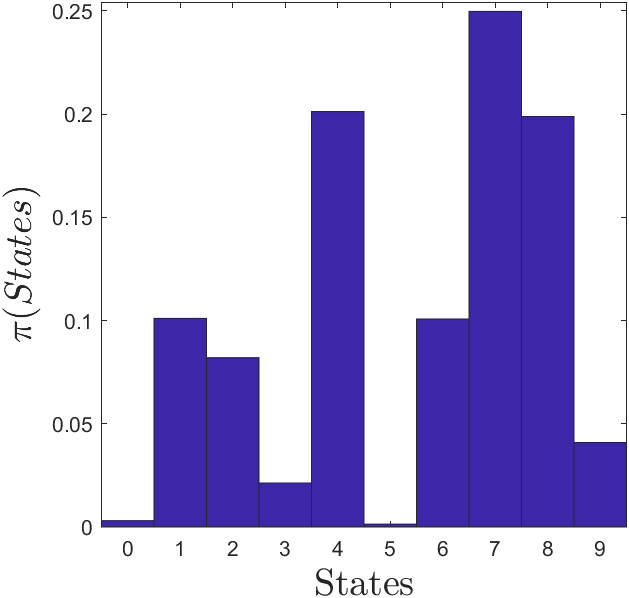} 
    \end{minipage}%
     \caption{Stationary distribution for the Network \ref{fig:BigLins}(b) with $k_{1} = 1$, $k_{2} = 2.5$, $k_{3} = 2$, $k_{4} = 1$, $k_{5}=1.75$, and $k_{6}=1.4$ and a state space that can be generated around the  point where there are $4$ molecules of $A$, and the remaining molecular counts equal to zero. The total variation norm distance between $\widehat{\pi}$ and $\pi$ was $0.004079$. \emph{\bf Left} Exact solution computed using \eqref{e:CME}. \emph{\bf Center} Perfectly simulated solution using $X \in \mathbb{S}$. \emph{\bf Right} Perfectly simulated solution using only $X \in \mathbb{S}_{C}$.}    

\vspace{0.15in}
    \begin{minipage}{.25\textwidth}
        \centering
        \includegraphics[width=\linewidth, height=0.2\textheight]{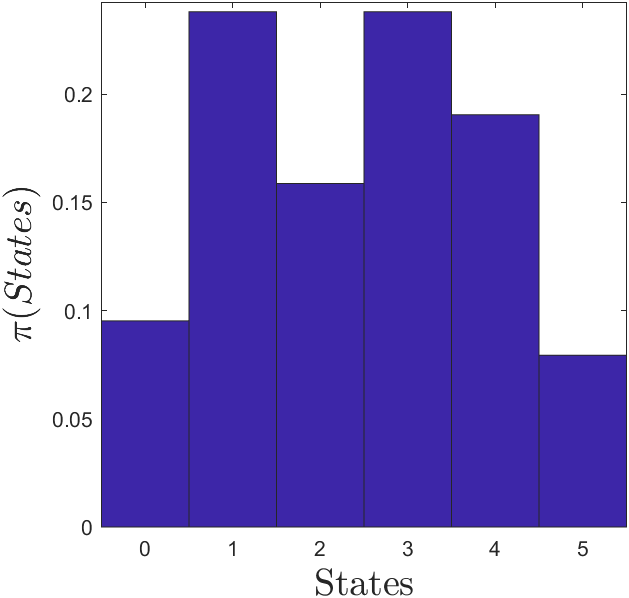} 
    \end{minipage}%
    \hfill
    \begin{minipage}{0.25\textwidth}
        \centering
        \includegraphics[width=\linewidth, height=0.2\textheight]{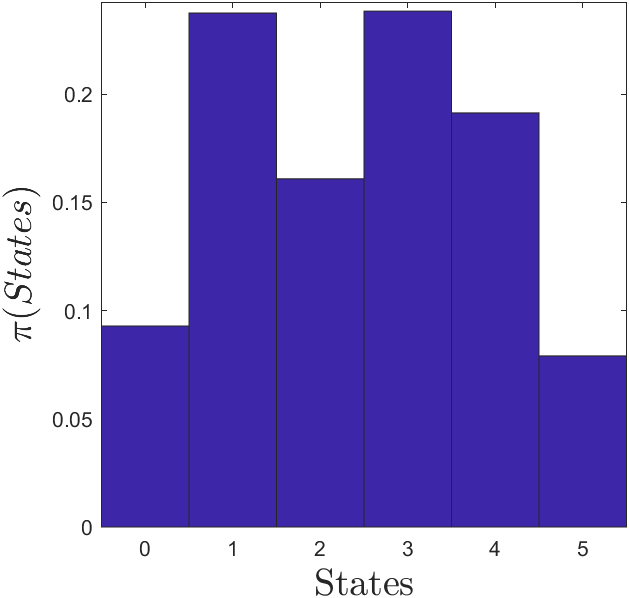} 
    \end{minipage}%
    \hfill
    \begin{minipage}{0.25\textwidth}
        \centering
        \includegraphics[width=\linewidth, height=0.2\textheight]{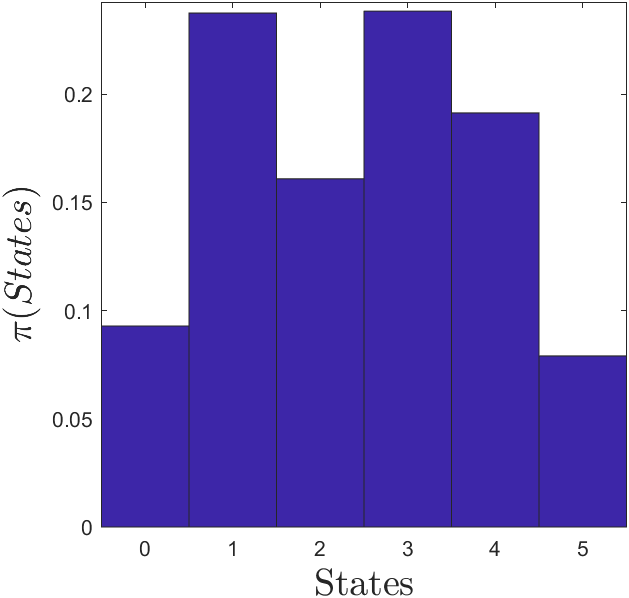} 
    \end{minipage}%
     \caption{Stationary distribution for the Network \ref{fig:BigLins}(c) with $k_{1} = 1$, $k_{2} = 2.5$, $k_{3} = 2$, $k_{4} = 1$, $k_{5}=1.75$, $k_{6}=1.4$, $k_{7}=0.25$, $k_{8}=0.375$, $k_{9}=0.9$, and $k_{10}=0.6$ and  a state space that can be generated around the  point where there are $3$ molecules of $A$, and the remaining molecular counts equal to zero. The total variation norm distance between $\widehat{\pi}$ and $\pi$ was $0.003318$.  \emph{\bf Left} Exact solution computed using \eqref{e:CME}. \emph{\bf Center} Perfectly simulated solution $X \in \mathbb{S}$. \emph{\bf Right} Perfectly simulated solution using only $X \in \mathbb{S}_{C}$.}  
     
    \label{fig:BigLins123}
\end{figure}

\begin{figure}[h]
    \centering
    \begin{minipage}{0.25\textwidth}
        \centering
        \includegraphics[width=\linewidth, height=0.2\textheight]{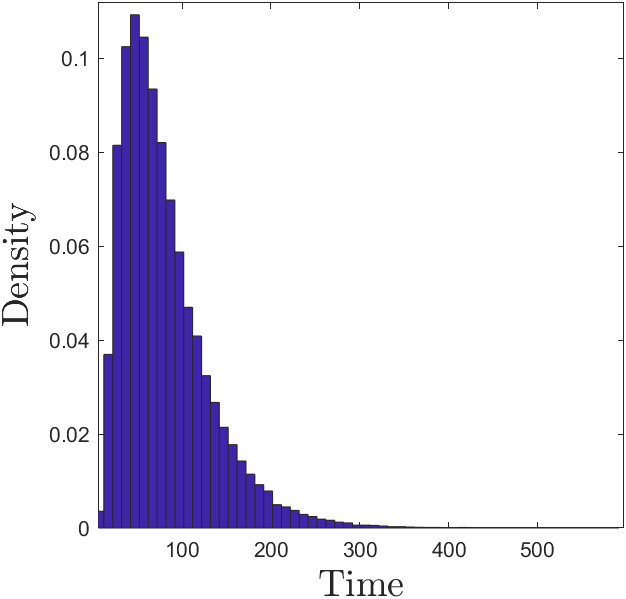} 
        \caption*{(a)}
    \end{minipage}%
    \hfill
    \begin{minipage}{0.25\textwidth}
        \centering
        \includegraphics[width=\linewidth, height=0.2\textheight]{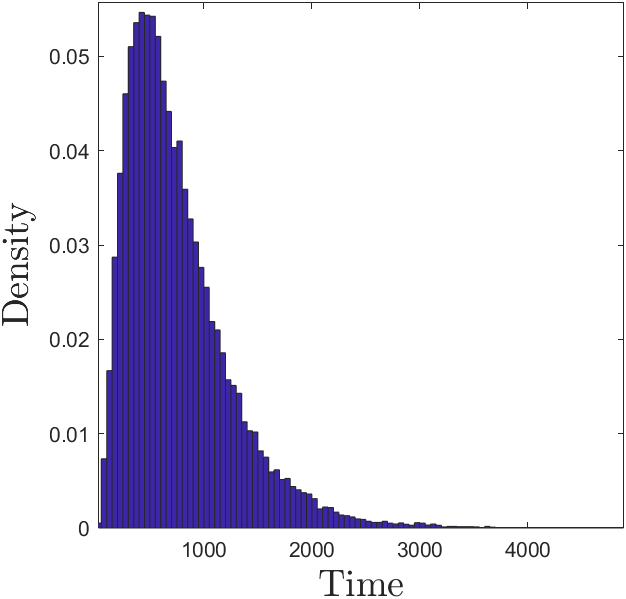} 
        \caption*{(b)}
    \end{minipage}%
    \hfill
    \begin{minipage}{0.25\textwidth}
        \centering
        \includegraphics[width=\linewidth, height=0.2\textheight]{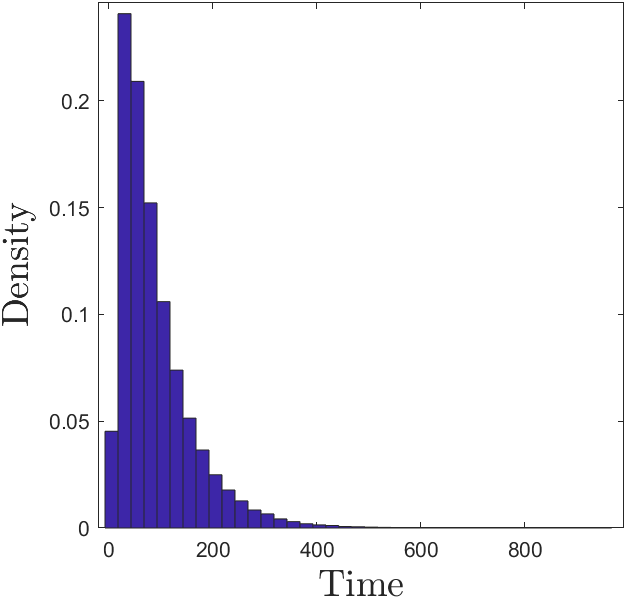} 
        \caption*{(c)}
    \end{minipage}%
    \hfill
    \begin{minipage}{0.25\textwidth}
        \centering
        \includegraphics[width=\linewidth, height=0.2\textheight]{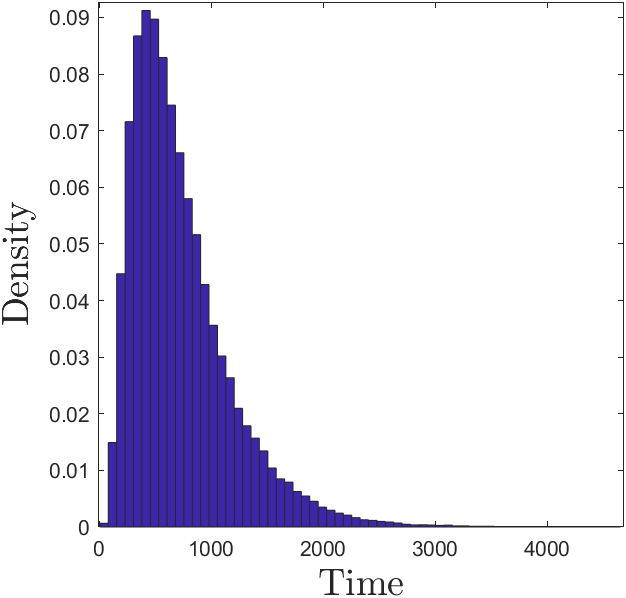} 
        \caption*{(d)}
    \end{minipage}%
\vspace{0.15in}      
    \begin{minipage}{0.25\textwidth}
        \centering
        \includegraphics[width=\linewidth, height=0.2\textheight]{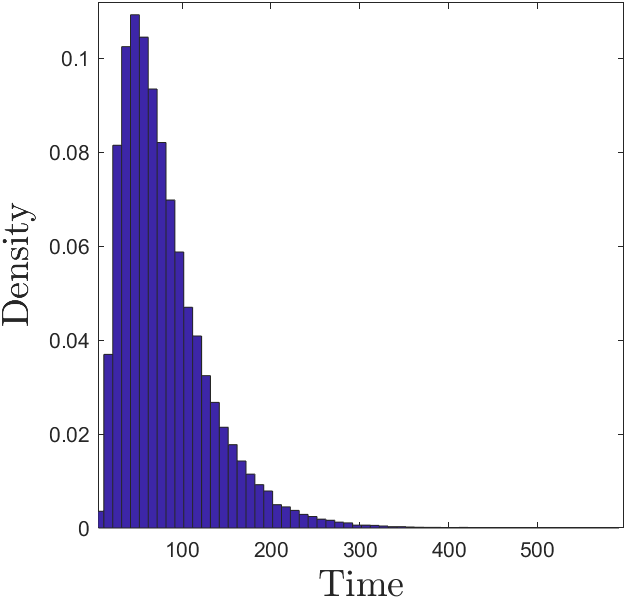} 
        \caption*{(a)}
    \end{minipage}%
    \hfill
    \begin{minipage}{0.25\textwidth}
        \centering
        \includegraphics[width=\linewidth, height=0.2\textheight]{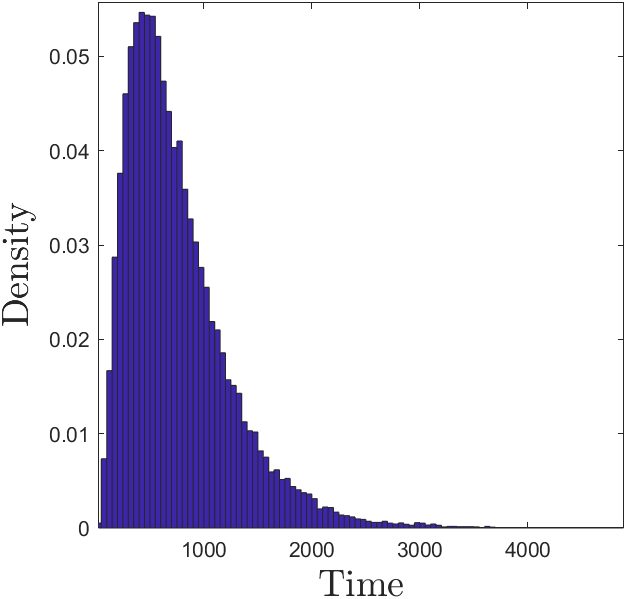} 
        \caption*{(b)}
    \end{minipage}%
    \hfill 
    \begin{minipage}{0.25\textwidth}
        \centering
        \includegraphics[width=\linewidth, height=0.2\textheight]{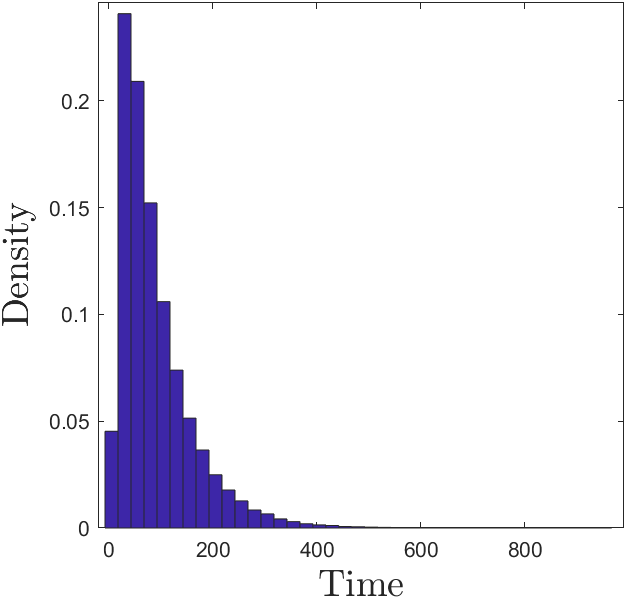} 
        \caption*{(c)}
    \end{minipage}%
    \begin{minipage}{0.25\textwidth}
        \centering
        \includegraphics[width=\linewidth, height=0.2\textheight]{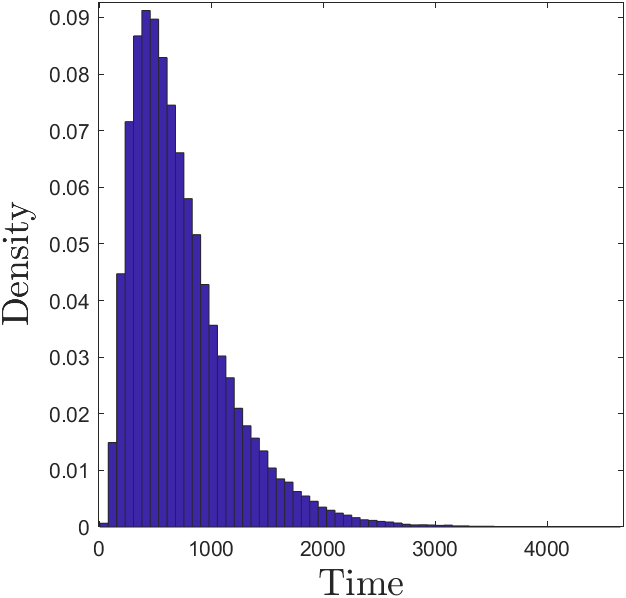} 
        \caption*{(d)}
    \end{minipage}%
     \caption{Distribution of backward coupling times for the networks in Figure \ref{fig:BigLins}. The \emph{\bf top row} shows coupling time distributions when paths are constructed for $X \in \mathbb{S}$ while the \emph{\bf bottom row} shows coupling time distributions when paths are constructed only for $X \in \mathbb{S}_{C}$. \emph{\bf Column 1} \ref{fig:BigLins}(a): In both simulations, the mean backward coupling time is $81.2079$, the median is $70$, and the mode is $45$. \emph{\bf Column 2} \ref{fig:BigLins}(b): In both simulations, the mean is $761.9452$, the median is $654$, and the mode is $345$. \emph{\bf Column 3} \ref{fig:BigLins}(c): In both distributions, the mean is $89.2597$, the median is $69$, and the mode is $32$. {\bf Column 4} \ref{fig:BigLins}(d): In both distributions, the mean is $728.7045$, the median is $621$, and the mode is $383$.}
     \label{fig:BigLinTimes123}
\end{figure}

The accuracy of the two simulation approaches, and the utility of the proposed subset $\mathbb{S}_{C}$, are underscored in the plotted distributions. In Figure \ref{fig:BigLins123} histograms of the simulated probability distributions for each network are compared to the exact distribution found using \eqref{e:CME}. To quantify the accuracy of each simulation, the total variation norm distance between the stationary distribution $\pi$ and the simulated distribution $\widehat{\pi}$ is reported. Figure \ref{fig:BigLinTimes123} shows the distributions of coupling times and their statistics which demonstrate the equivalent performance of the simulations utilizing only states in $\mathbb{S}_{C}$ and the simulations using all states in $\mathbb{S}$. We note that the statistics for the coupling time distributions are indistinguishable in each simulation. As the number of states increase, the use of this proposed subset allows the distribution of interest to be generated more efficiently than if all states are used and without sacrificing accuracy. 

\newpage 

\begin{figure}[!h]
    \centering
    \begin{minipage}{.3\textwidth}
        \centering
        \includegraphics[width=\linewidth, height=0.25\textheight]{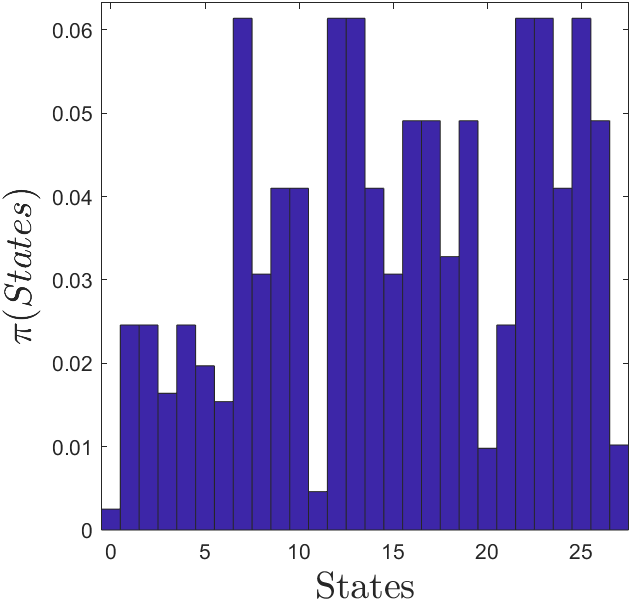} 
    \end{minipage}%
    \hfill
    \begin{minipage}{0.3\textwidth}
        \centering
        \includegraphics[width=\linewidth, height=0.25\textheight]{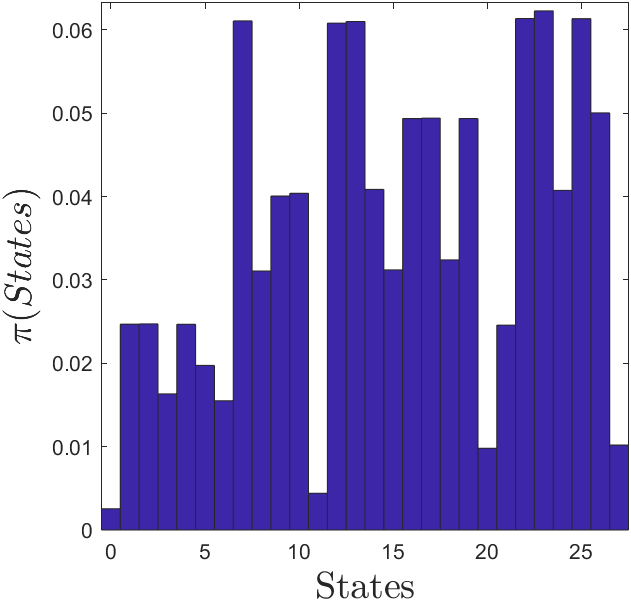} 
    \end{minipage}%
    \hfill
    \begin{minipage}{0.3\textwidth}
        \centering
        \includegraphics[width=\linewidth, height=0.25\textheight]{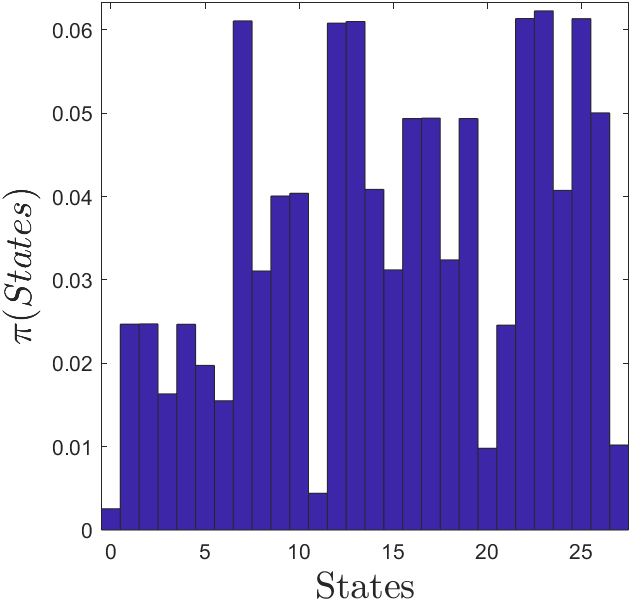} 
    \end{minipage}%
     \caption{Stationary distribution for the Network \ref{fig:BigLins}(d) with $k_{1} = 1$, $k_{2} = 2.5$, $k_{3} = 2$, $k_{4} = 1$, $k_{5}=1.75$, $k_{6}=1.4$, $k_{7}=0.25$, $k_{8}=0.375$, $k_{9}=0.9$, $k_{10}=0.6$, $k_{11}=1$, $k_{12}=1$ and a state space that can be generated around the point where there are 3 molecules of $A$ and the remaining molecular counts equal to zero. The total variation norm distance between $\widehat{\pi}$ and $\pi$ was $0.009530$.  \emph{\bf Left} Exact solution computed using \eqref{e:CME}. \emph{\bf Center} Perfectly simulated solution using $X \in \mathbb{S}$. \emph{\bf Right} Perfectly simulated solution using only $X \in \mathbb{S}_{C}$.}    
    \label{fig:BigLin4soln}
\end{figure}

\newpage 

\vspace{0.15in}

\newpage
\section{Conclusions}
\slabel{conclusions}

In this paper we provide a perfect-sampling algorithm for computing error-free draws from the stationary distributions of chemical reaction networks. This is in contrast to many popular sampling algorithms which make use of forward simulation and can only guarantee approximate draws from the distribution of interest. When the network is comprised of linear, reversible reactions between distinct chemical species, we proposed a subset of states that allows the distribution of interest to be computed more efficiently but without sacrificing accuracy, and we proved for a specific reaction network that this subset indeed guarantees coupling of all paths in the MCMC simulation. 

While our approach allows the distribution of interest to be computed accurately for a large class of networks, a clear obstacle to this method is the large average coupling times seen in the simulations of larger networks. This is likely due to poor mixing of the Markov chains resulting from low acceptance rates of moves proposed by the dominating maximum-rate process $\{Y(t)\}$. Numerous features influence the acceptance rate, such as the size of the state space and the form of the reaction rates used. A potential solution may be to treat acceptances as rare events and to use subset simulation techniques to increase the acceptance probability throughout the simulation. 

Additionally, there are many networks of interest that do not satisfy one (or more) of these properties. A natural next step is to apply this method to broader classes of reaction networks, such as cyclic reactions, irreversible reactions, and reactions composed of multiple linkage classes.

\section*{Appendix}

\begin{prop}{4.1}
    Let $\mathbb{S}$ be the state-space for the Reversible Michaelis-Menten chemical reaction network and let the coordinate pair $(x_{C},x_{P})$ represent a state in $\mathbb{S}$. Define the point $\mathcal{A}$ to be the state initially found at $(0,0)$ and the point $\mathcal{B}$ to be the state initially found at $(0,N)$. Evolve the reaction network forward according to reactions $R_{i}$ ($i = 1,2,3,4$).
    \begin{itemize} 
    \item
    \begin{example}{A}
        If $\mathcal{A} = (x_{C}^{\star},x_{P}^{\star})$ after $N$ reactions steps, then there are no points $(x_{C},x_{P}) \in \mathbb{S}$ such that $x_{P} < x_{P}^{\star}$ or $x_{C} < x_{C}^{\star}$, $x_{P} \leq -x_{C} + (x_{C}^{\star} + x_{P}^{\star})$
    \end{example}
    \item
    \begin{example}{B}
        If $\mathcal{B} = (x_{C}^{\star}, x_{P}^{\star})$ after $N$ reactions steps, then there are no points $(x_{C},x_{P}) \in \mathbb{S}$ such that $x_{P} > x_{P}^{\star}$ or $x_{C} > x_{C}^{\star}, \ x_{P} > -x_{C} + (x_{C}^{\star} + x_{P}^{\star})$.
    \end{example}
    \end{itemize}
\end{prop}

\subsection*{Statement and Proof of Claim A}
\begin{example}{A}
    Let $\mathbb{S}$ be the state-space for the Reversible Michaelis-Menten chemical reaction network and let the coordinate pair $(x_{C},x_{P})$ represent a state in $\mathbb{S}$. Define the point $\mathcal{A}$ to be the state initially found at $(x_{C},x_{P}) = (0,0)$. Evolve the reaction network forward according to reactions \genR ($i = 1,2,3,4$) so that $A = (x_{C}^{\star},x_{P}^{\star})$ after $k$ reactions steps. Then 
\begin{itemize}
    \item There are no points $(x_{C},x_{P})$ such that $x_{P} < x_{P}^{\star}$.
    \item There are no points $(x_{C},x_{P})$ such that $x_{C} < x_{C}^{\star}$, $x_{P} = x_{P}^{\star}$.
    \item There are no points $(x_{C},x_{P})$ such that $x_{C} < x_{C}^{\star}$, $x_{P} < -x_{C} + (x_{C}^{\star} + x_{P}^{\star})$.
\end{itemize}
\end{example} 

By induction:
\begin{itemize}
    \item Let $k=1$. The transitions rates from any state $(x_{C},x_{P})$ are
    \begin{align*}
        \lambda_{1}(x_{C},x_{P}) &= k_{1}(N-x_{C}-x_{P})(M-x_{C}) \\
        \lambda_{2}(x_{C},x_{P}) &= k_{2} \, x_{C} \\
        \lambda_{3}(x_{C},x_{P}) &= k_{3} \, x_{C} \\
        \lambda_{4}(x_{C},x_{P}) &= k_{4} \, x_{P} (M-x_{C})
    \end{align*}
    In particular, the one-step transition rates from $(0,0)$ are 
    \begin{align*}
        \lambda_{1}(0,0) = k_{1}NM, \quad 
        \lambda_{2}(0,0) = \lambda_{3}(0,0) = \lambda_{4}(0,0) = 0
    \end{align*}
    so $(x_{C}^{\star},x_{P}^{\star}) = (1,0)$ and $(0,0)$ is empty. Since $(0,0)$ is the only state below the line $x_{P} = -x_{C} + 1$, there is no occupied state such that $x_{P} = 0$ or $x_{P} < -x_{C} + 1$ when  $ x_{C} < 1$. \qed 
    
    \item Assume Claim A holds up through the first $k = n-1$ transitions. (``Case 1") \ Suppose that $\mathcal{A}$ moves during the $n-1 \mapsto n$ transition.
      \begin{itemize}
    	\setlength{\itemsep}{6pt} 
        \item[$\boldsymbol{R_1}$:] When $k=n-1$, the set of unoccupied states is
        \begin{align*}
            \big\{x_{P} < x_{P}^{\star} \big\} 
                \cup \big\{ x_{C} < x_{C}^{\star}-1, \ x_{P}=x_{P}^{\star} \big\} 
                \cup \big\{ x_{C} < x_{C}^{\star}-1, \ x_{P} < -x_{C} + (x_{C}^{\star} + x_{P}^{\star} - 1) \big\}
        \end{align*}
       After $n-1 \mapsto n$ via $R_1$, observe:
       \begin{itemize}
       \setlength{\itemsep}{3pt}
           \item[--] $R_1$ transitions move states horizontally to the right $(x_{C}.x_{P}) \mapsto (x_{C}+1,x_{P})$. Thus, any state satisfying $x_{P} < x_{P}^{\star}$ that undergoes an $R_1$ transition will move to another state satisfying $x_{P} < x_{P}^{\star}$, while any state satisfying $x_{P} > x_{P}^{\star}$ will move to another state satisfying $x_{P} > x_{P}^{\star}$. $\big\{x_{P} < x_{P}^{\star} \big\}$ is still empty since $\{x_{P} \ (k=n-1)\} = \{x_{P} \ (k=n)\} = x_{P}^{\star}$.
             
           \item[--] The $R_1$ transition rate of $(x_{C}^{\star}-1,x_{P}^{\star})$ is 
            \begin{align*}
                \lambda_{1}(x_{C}^{\star}-1,x_{P}^{\star}) = k_{1} (N + 1 - x_{C}^{\star} - x_{P}^{\star}) (M + 1 - x_{C}^{\star})
            \end{align*}
            Every point $(x_{C},x_{P})$ on the line $x_{P} = -x_{C} + (x_{C}^{\star} + x_{P}^{\star} - 1)$ satisfies $x_{C} = (x_{C}^{\star}-1) - \alpha$, $x_{P} = x_{P}^{\star} + \alpha$ and has the $R_1$ transition rate
            \begin{align*}
                \lambda_{1}(x_{C},x_{P}) = k_{1} (N + 1 - x_{C}^{\star} - x_{P}^{\star}) (M + 1 + \alpha - x_{C}^{\star})
            \end{align*}
            If $x_{C} < x_{C}^{\star}-1$, then $\lambda_{1}(x_{C},x_{P}) > \lambda_{1}(x_{C}^{\star}-1,x_{P}^{\star})$ and these points on the line must move with $\mathcal{A}$ during the $n-1 \mapsto n$ transition. There are no points to the left of this line to occupy the vacated spaces (inductive assumption), so the line $x_{P} = -x_{C} + (x_{C}^{\star} + x_{P}^{\star} -1)$ is empty if $x_{C} < x_{C}^{\star}$. 
           
         \item[--] Since $\ds \big\{ x_{P}=x_{P}^{\star}, \ x_{C} < x_{C}^{\star} - 1 \big\} \subset \big\{ x_{P}=x_{P}^{\star}, \ x_{C} < x_{C}^{\star} \big\} $ is not filled in during an $R_1$ move, we only need to check the point $(x_{C}^{\star}  - 1, x_{P}^{\star})$. But this was the position of $\mathcal{A}$ prior to the $n-1 \mapsto n$ transition, so it must also be empty once $k=n$.     \qed
       \end{itemize}
       $\implies$ Claim A holds after an $R_1$ transition. \qed
            
        \item[$\boldsymbol{R_2}$:]  When $k=n-1$, the set of unoccupied states is
        \begin{align*}
            \big\{x_{P} < x_{P}^{\star} \big\} 
                \cup \big\{ x_{C} < x_{C}^{\star}+1, \ x_{P}=x_{P}^{\star} \big\} 
                \cup \big\{ x_{C} < x_{C}^{\star} + 1, \ x_{P} < -x_{C} + (x_{C}^{\star} + x_{P}^{\star} + 1) \big\}
        \end{align*}
       After $n-1 \mapsto n$ via $R_2$, observe:
         \begin{itemize}
         \setlength{\itemsep}{3pt}
             \item[--] $R_2$ transitions move states horizontally to the left $(x_{C},x_{P}) \mapsto (x_{C}-1,x_{P})$. Thus, any state satisfying $x_{P} < x_{P}^{\star}$ that undergoes an $R_2$ transition will move to another state satisfying $x_{P} < x_{P}^{\star}$, while any state satisfying $x_{P} > x_{P}^{\star}$ will move to another state satisfying $x_{P} > x_{P}^{\star}$. $\big\{x_{P} < x_{P}^{\star} \big\}$ is still empty since $\{x_{P} \ (k=n-1) \} = \{x_{P} \ (k=n)\} = x_{P}^{\star}$.
             
             \item[--] $\big\{ x_{C} < x_{C}^{\star}+1, \ x_{P} = x_{P}^{\star} \big\} = \big\{ x_{C} < x_{C}^{\star}, x_{P} = x_{P}^{\star} \big\} \cup \big\{ x_{C} = x_{C}^{\star}, \ x_{P} = x_{P}^{\star} \big\}$. After the $R_2$ transition, $\big\{ x_{P} = x_{P}^{\star}, \ x_{C} < x_{C}^{\star} \big\}$ is still empty.
             
             \item[--] $\big\{ x_{C} < x_{C}^{\star} + 1, \ x_{P} < -x_{C} + (x_{C}^{\star} + x_{P}^{\star} + 1)\big\} = \big\{ x_{C} < x_{C}^{\star}, \ x_{P} < -x_{C} + (x_{C}^{\star} + x_{P}^{\star})\big\} \cup \big\{ x_{C} \leq x_{C}^{\star}, \ x_{P} = -x_{C} + (x_{C}^{\star} + x_{P}^{\star})\big\}$. After the $R_2$ transition, $ \big\{ x_{C} < x_{C}^{\star}, \ x_{P} < -x_{C} + (x_{C}^{\star} + x_{P}^{\star}) \big\}$ is still empty.
         \end{itemize}
         $\implies$ Claim A holds after an $R_2$ transition. \qed
         
        \item[$\boldsymbol{R_3}$:]  When $k=n-1$, the set of unoccupied states is
        \begin{align*}
            \big\{ x_{P} < x_{P}^{\star} - 1 \big\}
                \cup \big\{ x_{C} < x_{C}^{\star} + 1, \ x_{P}=x_{P}^{\star}-1 \big\}
                \cup \big\{ x_{C} < x_{C}^{\star} + 1, \ x_{P} < -x_{C} + (x_{C}^{\star} + x_{P}^{\star}) \big\}
        \end{align*}
        After $n-1 \mapsto n$ via $R_3$, observe:
        \begin{itemize}
        \setlength{\itemsep}{3pt}
            \item[--] The states satisfying $x_{P} < x_{P}^{\star}-1$ are still empty. Furthermore, the transition rates of all states undergoing an $R_3$ move in the $n$-th reaction step satisfy $\ds k_{3}(x_{C}^{\star} + 1) < k_{3} \, x_{C}$ for $x_{C} > x_{C}^{\star}$. In particular, the states
                \begin{align*}
                     (x_{C}^{\star} + 1, x_{P}^{\star} -1), \ (x_{C}^{\star} + 2, x_{P}^{\star} - 1), \ (x_{C}^{\star} + 3, x_{P}^{\star} - 1), \hdots (M, x_{P}^{\star} -1)
                \end{align*}
            satisfy this inequality and must be vacated in the $n$-th reaction step. So $x_{P} \leq x_{P}^{\star} - 1 < x_{P}^{\star}$ must be empty after the $R_3$ transition.
            
            \item[--] $\big\{ x_{C} < x_{C}^{\star}+1, \ x_{P} < -x_{C} + (x_{C}^{\star} + x_{P}^{\star})\big\} = \big\{ x_{C} < x_{C}^{\star}, \ x_{P} < -x_{C} + (x_{C}^{\star} + x_{P}^{\star}) \big\} \cup \big\{ x_{C} = x_{C}^{\star}, \ x_{P} < -x_{C} + (x_{C}^{\star} + x_{P}^{\star})\big\}$. After the $R_3$ transition, $\big\{ x_{C} < x_{C}^{\star}, \ x_{P} < -x_{C} + (x_{C}^{\star} + x_{P}^{\star}) \big\}$ is still empty.
            
            
            \item[--] From the above argument, $(x_{C},x_{P})$ satisfying $x_{P} < -x_{C} + (x_{C}^{\star} + x_{P}^{\star})$ and $x_{C} < x_{C}^{\star}$ are empty. In particular, they are empty for $x_{P} = x_{P}^{\star}$. 
        \end{itemize}
\vspace{0.1in}        
        \item[$\boldsymbol{R_4}$:]  When $k=n-1$, the set of unoccupied states is
        \begin{align*}
            \big\{ x_{P} < x_{P}^{\star} + 1 \big\}
                \cup \big\{ x_{C} < x_{C}^{\star} - 1, \ x_{P}=x_{P}^{\star}+1 \big\}
                \cup \big\{ x_{C} < x_{C}^{\star} - 1, \ x_{P} < -x_{C} + (x_{C}^{\star} + x_{P}^{\star}) \big\}
        \end{align*}
        After $n-1 \mapsto n$ via $R_4$, observe:
        \begin{itemize}
        \setlength{\itemsep}{4pt}        
            \item[--] The states $(x_{C},x_{P}) \in \big\{x_{C} < x_{C}^{\star} -1 , \ x_{P} \geq x_{P}^{\star} + 1 \big\}$ satisfy $k_{4} \, x_{P} (M-x_{C}) > k_{4} (x_{P}^{\star} + 1) (M - (x_{C}^{\star}-1))$ and must make an $R_4$ transition. The states within the triangle $\big\{ x_{C} < x_{C}^{\star}-1, \ x_{P}^{\star}+1 \leq x_{P} < -x_{C} + (x_{C}^{\star} + x_{P}^{\star}) \big\}$ are already empty (IA) and the states on the line $x_{P} = -x_{C} + (x_{C}^{\star} + x_{P}^{\star})$ will remain on the line during the $R_4$ move, so $\big\{ x_{C} < x_{C}^{\star}-1, \ x_{P}^{\star}+1 \leq x_{P} < -x_{C} + (x_{C}^{\star} + x_{P}^{\star}) \big\}$ is still empty.
            
            \item[--] The states $\big\{x_{P} = x_{P}^{\star}+1, x_{C} < x_{C}^{\star}-1 \big\}$ must move but are already empty. Since also $\big\{ x_{P} < x_{P}^{\star} + 1 \big\}$ was empty for $k=n-1$ (IA) then the states $\big\{ x_{P} = x_{P}^{\star}, x_{C} < x_{C}^{\star}-1\big\}$ remain empty. To conclude that $\big\{ x_{P} = x_{P}^{\star}, \ x_{C} < x_{C}^{\star}\big\}$ is empty, we must simply check the state $(x_{C}^{\star}-1,x_{P}^{\star})$. Note $(x_{C}^{\star}-1,x_{P}^{\star})$ was empty prior to the transition. The $R_4$ move caused $(x_{C}^{\star}-2,x_{P}^{\star}+1) \mapsto (x_{C}^{\star}-1,x_{P}^{\star})$. Since $(x_{C}^{\star}-2,x_{P}^{\star}+1) \in \big\{ x_{P} = x_{P}^{\star}+1, x_{C} < x_{C}^{\star}-1\big\}$ then this state was also empty before the move, so the state $(x_{C}^{\star}-1,x_{P}^{\star})$ remains empty.
            
            \item[--]  $ \big\{ x_{P} < x_{P}^{\star} + 1\big\} =  \big\{ x_{P} < x_{P}^{\star} \big\} \cup \big\{ x_{P} = x_{P}^{\star} \big\}$ was empty. All states along the line $x_{P} = x_{P}^{\star}+1$ may move in the $R_4$ transition. Those satisfying $x_{C} < x_{C}^{\star}-1$ must necessarily move, but are empty, and this move will not affect the empty states in $\big\{x_{C} < x_{C}^{\star}-1, \ x_{P} = x_{P}^{\star}\big\}$. OTOH, the occupied states $\big\{ x_{C} \geq x_{C}^{\star}, \ x_{P} = x_{P}^{\star}+1 \big\}$ may move and fill $\big\{ x_{C} > x_{C}^{\star}, \ x_{P} = x_{P}^{\star} \big\}$. Since $R_4$ is a one-step transition, then none of the states along the line $x_{P} = x_{P}^{\star} - 1$ or below are filled. $\big\{ x_{P} < x_{P}^{\star} \big\}$ remains empty. 
        \end{itemize}
    $\implies$ Claim A holds after an $R_4$ transition. \qed
    \end{itemize}
    
\vspace{0.1in}     
    \item Assume Claim A holds for $k=n-1$ transitions. (``Case 2") \ Suppose that $A = (x_{C}^{\star},x_{P}^{\star})$ when $k=n$ but $\mathcal{A}$ did not necessarily move during the $n-1 \mapsto n$ reaction step. 
    \begin{itemize}
    \setlength{\itemsep}{6pt}
        \item See ``Case 1" in the event $\mathcal{A}$ moves during the $n-1 \mapsto n$ transition.
        \item[] If state $\mathcal{A}$ is at $(c,p)$ when $k=n-1$, define the necessarily empty sets with
        \begin{align*}
            \mathcal{R} &= \big\{x_{P} < p \big\} \\
            \mathcal{T} &= \big\{ x_{C} < c, \ p \leq x_{P} < - x_{C}+(c + p)\big\}
        \end{align*}
        \item[$\boldsymbol{R_1}$:] Note: $\mathcal{R}$ is empty for $k=n-1$ (IA). $R_1$ and $R_2$ transitions involve purely right and left moves, respectively. Any state in $\mathcal{R}$ that undergoes an $R_1$ or $R_2$ transition will move to another state in $\mathcal{R}$, while any state not in $\mathcal{R}$ that undergoes an $R_1$ or $R_2$ transition will move to another state not in $\mathcal{R}$. From this we know $\mathcal{R}$ will remain empty under $R_1$ and $R_2$ transitions and we simply check what happens to $\mathcal{T}$ during these transitions.\\
        \\
        States may move ``right" in the state space through an $R_1$ transition and we must check if a state will violate the assertion that $\mathcal{T}$ is empty. If state $\mathcal{A}$ is at $(c,p)$ when $k=n-1$, then a state $(x_{C},x_{P})$ satisfying $x_{P} \geq - x_{C} + (c + p)$ will move away from $\mathcal{T}$ during an $R_1$ transition, leaving $\mathcal{T}$ empty. Every state $(x_{C},x_{P}) \in \mathcal{T}$ is unoccupied for $k=n-1$ and an $R_1$ move will map empty states to neighboring empty states.  $\mathcal{T}$ remains empty. \qed
        
        \item[$\boldsymbol{R_2}$:] (See above for effect of $R_2$ move on states in $\mathcal{R}$). States may move ``left" in the state space through an $R_2$ transitions, so we must check if a state will violate the assertion $\mathcal{T}$ is empty. If state $\mathcal{A}$ is at $(c,p)$ when $k=n-1$, all of the (possibly) nonempty candidates to move left into the triangle $\mathcal{T}$ in a single transition step lie along the line $x_{P} = - x_{C} + (c + p)$ and can be represented by $(c - \alpha, p + \alpha)$ where $\alpha > 0$. The $R_2$ transition rates for these states are 
        \begin{align*}
            \lambda_{2}(x_{C},x_{P}) = k_{2} (c - \alpha)
                     < k_{2} \, c
                     = \lambda_{2}(c,p)
        \end{align*}
        Therefore, if a state on $x_{P} = - x_{C} + (c+p)$ moves during an $R_2$ transition step, then state $\mathcal{A}$ must necessarily undergo an $R_2$ transition also. $\mathcal{T}$ remains empty. \qed
        
        \item[$\boldsymbol{R_3}$:] States may move ``up" in the state space through an $R_3$ transition, so we must check if a state will violate the assertion $\mathcal{T}$ is empty. All of the (possibly) nonempty candidates to move left into the triangle $\mathcal{T}$ in a single transition step lie along the line $x_{P} = - x_{C} + (c + p)$ and can be represented by $(c - \alpha, p + \alpha)$. However, every state undergoing an $R_3$ move will transition along a diagonal line. In particular, the candidate states will move along the line $x_{P} = - x_{C} + (c + p)$ and will not enter $\mathcal{T}$. A state that lines along $x_{P} = p-1$ with $x_{C} < c$ may enter $\mathcal{T}$ via an $R_3$ move, but these states are in $\mathcal{R}$ and already empty (IA). $\mathcal{T}$ remains empty. \qed
        
        \item[$\boldsymbol{R_4}$:] States may move ``down" in the state space through an $R_4$ transition, so we must check if a state will violate the assertion that $\mathcal{R}$ is empty. (Note: any point undergoing an $R_4$ move will also move to the right during this transition.) All of the candidates to move below $\mathcal{A}$ in a single transition step must occupy states along $\big\{x_{P} = p \big\}$. The states with $x_{P}=p$ and $x_{C}<c$ are empty by the induction assumption. If $x_{C} > c$, then the $R_4$ transition rates are proportional to $\lambda_{4}(x_{C},p) = k_{4} \, p (M - x_{C}) < k_{4} \, p (M - c) = \lambda_{4}(c,p)$. If these states undergo $R_4$ transitions, then $\mathcal{A}$ must necessarily move through an $R_4$ transition also. $\mathcal{R}$ remains empty. \qed

    \end{itemize}
\end{itemize}


\newpage 
\subsection*{Statement and Proof of Claim B}
\begin{example}{B}
    Let $\mathbb{S}$ be the state-space for the Reversible Michaelis-Menten chemical reaction network and let the coordinate pair $(x_{C},x_{P})$ represent a state in $\mathbb{S}$. Define the point $\mathcal{B}$ to be the state initially found at $(x_{C},x_{P}) = (0,N)$. Evolve the reaction network forward according to reactions \genR ($i = 1,2,3,4$) so that $B = (x_{C}^{\star},x_{P}^{\star})$ after $k$ reactions steps. Then 
\begin{itemize}
    \item There are no points $(x_{C},x_{P})$ such that $x_{P} > x_{P}^{\star}$.
    \item There are no points $(x_{C},x_{P})$ such that $x_{C} > x_{C}^{\star}, \ x_{P} > -x_{C} + (x_{C}^{\star} + x_{P}^{\star})$.
\end{itemize}
\end{example} 

    By induction:
\begin{itemize}
    \item Let $k=1$. The transitions rates from any state $(x_{C},x_{P})$ are
    \begin{align*}
        \lambda_{1}(x_{C},x_{P}) &= k_{1}(N-x_{C}-x_{P})(M-x_{C}) \\
        \lambda_{2}(x_{C},x_{P}) &= k_{2} \, x_{C} \\
        \lambda_{3}(x_{C},x_{P}) &= k_{3} \, x_{C} \\
        \lambda_{4}(x_{C},x_{P}) &= k_{4} \, x_{P} (M-x_{C})
    \end{align*}
    In particular, the one-step transition rates from $(0,N)$ are 
    \begin{align*}
        \lambda_{1}(0,0) = \lambda_{2}(0,0) = \lambda_{3}(0,0) = 0, \quad
        \lambda_{4}(0,N) = k_{4}NM
    \end{align*}
    so $(x_{C}^{\star},x_{P}^{\star}) = (1,N-1)$ and $(0,N)$ is empty. Since $(0,N)$ is the only state satisfying $x_{P} > N-1 $, there is no occupied state such that $x_{P} > N-1$ or $x_{C} > 1, \ x_{P} > -x_{C} + N $. \qed 
    
    
    \item Assume Claim B holds up through the first $k = n-1$ transitions. (``Case 1") \ Suppose that $\mathcal{B}$ moves during the $n-1 \mapsto n$ transition.
      \begin{itemize}
    	\setlength{\itemsep}{6pt} 
        \item[$\boldsymbol{R_1}$:] When $k=n-1$, the set of unoccupied states is
        \begin{align*}
            \big\{x_{P} > x_{P}^{\star} \big\} 
                \cup \big\{x_{C} > x_{C}^{\star}-1, \ x_{P} > -x_{C} + (x_{C}^{\star} + x_{P}^{\star} - 1) \big\}
        \end{align*}
       After $n-1 \mapsto n$ via $R_1$, observe:
       \begin{itemize}
       \setlength{\itemsep}{5pt}
           \item[--] $R_1$ transitions move states horizontally to the right $(x_{C},x_{P}) \mapsto (x_{C}+1,x_{P})$. Thus, any state satisfying $x_{P} > x_{P}^{\star}$ that undergoes an $R_1$ transition will move to another state satisfying $x_{P} > x_{P}^{\star}$, while any state satisfying $x_{P} < x_{P}^{\star}$ will move to another state satisfying $x_{P} < x_{P}^{\star}$. $\big\{ x_{P} > x_{P}^{\star} \big\}$ is still empty since $\{ x_{P} \ (k=n-1) \} = \{x_{P} \ (k=n)\} = x_{P}^{\star}$. 
           
           \item[--] Note $\big\{ x_{C} > x_{C}^{\star}-1, \ x_{P} > -x_{C} + (x_{C}^{\star} + x_{P}^{\star} -1) \big\} = \big\{ x_{C} \geq x_{C}^{\star}, \ x_{P}  = - x_{C} + (x_{C}^{\star} + x_{P}^{\star}) \big\} \cup \big\{ x_{C} > x_{C}^{\star}, \ x_{P} > -x_{C} + (x_{C}^{\star} + x_{P}^{\star} ) \big\}$. After the $R_1$ transition, $\big\{ x_{C} > x_{C}^{\star}, \ x_{P} > -x_{C} + (x_{C}^{\star} + x_{P}^{\star})\big\}$ is still empty.
       \end{itemize}
       $\implies$ Claim B holds after an $R_1$ transition. \qed
       
       \item[$\boldsymbol{R_2}$:] When $k=n-1$, the set of unoccupied states is
       \begin{align*}
           \big\{x_{P} > x_{P}^{\star} \big\} 
                \cup \big\{x_{C} > x_{C}^{\star}+1, \ x_{P} > -x_{C} + (x_{C}^{\star} + x_{P}^{\star} + 1) \big\}
       \end{align*}
       After $n-1 \mapsto n$ via $R_2$, observe:
       \begin{itemize}
           \item[--] $R_2$ transitions move states horizontally to the left $(x_{C},x_{P}) \mapsto (x_{C}-1,x_{P})$. Thus, any state satisfying $x_{P} > x_{P}^{\star}$ that undergoes an $R_2$ transition will move to another state satisfying $x_{P} > x_{P}^{\star}$, while any state satisfying $x_{P} < x_{P}^{\star}$ will move to another state satisfying $x_{P} < x_{P}^{\star}$. $\big\{ x_{P} > x_{P}^{\star} \big\}$ is still empty since $\{ x_{P} \ (k=n-1)\} = \{ x_{P} \ (k=n)\} = x_{P}^{\star}$. 
           
           \item[--] Every point $(x_{C},x_{P})$ on the line $x_{P} = -x_{C} + (x_{C}^{\star} + x_{P}^{\star} + 1)$ satisfies $x_{C} = (x_{C}^{\star} + 1) + \alpha$, $x_{P} = x_{P}^{\star} - \alpha$ and has the $R_2$ transition rate
           \begin{align*}
               \lambda_{2}(x_{C},x_{P}) &= k_{2} (x_{C}^{\star} + 1 + \alpha)
           \end{align*}
           If $x_{C} > x_{C}^{\star}+1$ then $\lambda_{2}(x_{C},x_{P}) > \lambda_{2}(x_{C}^{\star}+1,x_{P}^{\star})$ and these points on the line must move with $\mathcal{B}$ during the $n-1 \mapsto n$ transition. There are no points to the right of the line to occupy the vacated space (induction assumption), so the line $x_{P} = -x_{C} + (x_{C}^{\star} + x_{P}^{\star} + 1)$ is empty if $x_{C} > x_{C}^{\star}$.
       \end{itemize}
       $\implies$ Claim B holds after an $R_2$ transition. \qed

        \item[$\boldsymbol{R_3}$:] When $k=n-1$, the set of unoccupied states is
        \begin{align*}
            \big\{x_{P} > x_{P}^{\star} - 1\big\} 
                \cup \big\{ x_{C} > x_{C}^{\star} + 1, \ x_{P} > -x_{C} + (x_{C}^{\star} + x_{P}^{\star}) \big\}
        \end{align*}
        After $n-1 \mapsto n$ via $R_3$, observe:
        \begin{itemize}
            \item[--] Note $\big\{ x_{P} > x_{P}^{\star} - 1 \big\} = \big\{ x_{P} = x_{P}^{\star} \big\} \cup \big\{ x_{P} > x_{P}^{\star} \big\}$ was empty. All states along the line $x_{P} = x_{P}^{\star} - 1$ may move in the $R_3$ transition. Those states with $x_{C} > x_{C}^{\star} + 1$ must necessarily move, since the transition rates for the $R_3$ transition will satisfy $\lambda_{3}(x_{C},x_{P}) > \lambda_{3}(x_{C}^{\star}+1,x_{P}^{\star})$, and may fill empty states along the line $x_{P} = x_{P}^{\star}$. However, the transition advances only one step and none of the states above the line $x_{P} = x_{P}^{\star}$ will be filled. $\big\{ x_{P} = x_{P}^{\star} \big\}$ will remain empty.
            
            \item[--] The states $(x_{C},x_{P}) \in \big\{ x_{C} > x_{C}^{\star} + 1, \ x_{P} > -x_{C} + (x_{C}^{\star} + x_{P}^{\star})$ satisfy $ k_{3} \, x_{C} > k_{3} \, (x_{C}^{\star} + 1)$ and must move with $\mathcal{B}$. Since $x_{P} > -x_{C} + (x_{C}^{\star} + x_{P}^{\star})$ was empty for $x_{C} > x_{C}^{\star} + 1$ (IA), and since the states on the lines $x_{P} = -x_{C} + (x_{C}^{\star} + x_{P}^{\star})$ will remain on the line during the $R_3$ move, then $\big\{ x_{C} > x_{C}^{\star}, \ x_{P} > -x_{C} + (x_{C}^{\star} + x_{P}^{\star})\big\}$ will still be empty.
        \end{itemize}
        $\implies$ Claim B holds after an $R_3$ transition. \qed

        \item[$\boldsymbol{R_4}$:] When $k = n-1$, the set of unoccupied states is
        \begin{align*}
            \big\{ x_{P} > x_{P}^{\star} + 1 \big\} 
                \cup \big\{ x_{C} > x_{C}^{\star} - 1, \ x_{P} > -x_{C} + (x_{C}^{\star} + x_{P}^{\star} )\big\}
        \end{align*}
        After $n - 1 \mapsto n$ via $R_4$, observe:
        \begin{itemize}
            \item[--] If states $(x_{C},x_{P})$ are on the line $x_{P} = x_{P}^{\star} + 1$ and if $x_{C} < x_{C}^{\star}-1$, then the $R_4$ transition rates for these states satisfy $k_{4} (x_{P}^{\star}+1)(M-x_{C}) > k_{4} (x_{P}^{\star}+1)(M + 1 -x_{C}^{\star})$ and these states must move with $\mathcal{B}$. The states on the line $\big\{ x_{C} < x_{C}^{\star} - 1, \ x_{P} = x_{P}^{\star} + 1 \big\}$ will necessarily be left empty. Furthermore, if $x_{C} > x_{C}^{\star} - 1$ then by the induction assumption the states $\big\{ x_{C} > x_{C}^{\star}-1, \ x_{P} > -x_{C} + (x_{C}^{\star} + x_{P}^{\star})\big\}$ are empty. In particular, the points on $\big\{ x_{C} > x_{C}^{\star}-1, \ x_{P} = x_{P}^{\star} + 1 \big\}$ are empty and will remain empty after the $R_4$ move. Finally, $(x_{C}^{\star}-1,x_{P}^{\star}+1)$ is the position of $\mathcal{B}$ when $k=n-1$ and this state will be left empty after the $R_4$ transition. $\big\{ x_{P} > x_{P}^{\star}\big\}$ will remain empty.
            
            \item[--] $\big\{ x_{C} > x_{C}^{\star} - 1, \ x_{P} > -x_{C} + (x_{C}^{\star} + x_{P}^{\star}) \big\} = \big\{ x_{C} = x_{C}^{\star}, x_{P} > -x_{C} + (x_{C}^{\star} + x_{P}^{\star})\big\} \cup \big\{ x_{C} > x_{C}^{\star}, \ x_{P} > -x_{C} + (x_{C}^{\star} + x_{P}^{\star})\big\}$ is empty when $k = n-1$. $R_4$ transitions move states diagonally along the lines $(x_{C}-1,x_{P}+1) \mapsto (x_{C},x_{P})$. In particular, the states on the line $x_{P} = -x_{C} + (x_{C}^{\star} + x_{P}^{\star})$ will remain on this line under the $R_4$, while any states $(x_{C},x_{P})$ satisfying $x_{P} > -x_{C} + (x_{C}^{\star} + x_{P}^{\star})$ when $x_{C} > x_{C}^{\star} - 1$ are empty (induction assumption) and will be mapped to other empty states satisfying the same inequality. $\big\{ x_{C} > x_{C}^{\star}, \ x_{P} > -x_{C} + (x_{C}^{\star} + x_{P}^{\star})\big\}$ remains empty.        
        \end{itemize}
        $\implies$ Claim B holds after an $R_4$ transition. \qed
    \end{itemize}   
       
     \item Assume Claim B holds for $k=n-1$ transitions. (``Case 2") \ Suppose that $B = (x_{C}^{\star},x_{P}^{\star})$ when $k=n$ but $\mathcal{B}$ did not necessarily move during the $n-1 \mapsto n$ reaction step. 
    \begin{itemize}
    \setlength{\itemsep}{6pt}
        \item See ``Case 1" in the event $\mathcal{B}$ moves during the $n-1 \mapsto n$ transition.
        \item[] If state $\mathcal{B}$ is at $(c,p)$ when $k=n-1$, define the necessarily empty sets with
        \begin{align*}
            \mathcal{T} &= \big\{x_{P} > p \big\} \\
            \mathcal{L} &= \big\{ x_{C} > c, \ x_{P} > -x_{C}+(c + p)\big\}
        \end{align*}
       \item[$\boldsymbol{R_1}$:] Note: $\mathcal{T}$ is empty for $k=n-1$ (IA). $R_1$ and $R_2$ transitions involve purely right and left moves, respectively. Any state in $\mathcal{T}$ that undergoes an $R_1$ or $R_2$ transition will move to another state in $\mathcal{T}$, while any state not in $\mathcal{T}$ that undergoes an $R_1$ or $R_2$ transition will move to another state not in $\mathcal{T}$. From this we know $\mathcal{T}$ will remain empty under $R_1$ and $R_2$ transitions and we simply check what happens to $\mathcal{L}$ during these transitions.\\
        \\
        States may move ``right" in the state space through an $R_1$ transition and we must check if a state will violate the assertion that $\mathcal{L}$ is empty. If state $\mathcal{B}$ is at $(c,p)$ when $k=n-1$, then the candidate states to occupy the vacant positions $\big\{ x_{C} > c, \ x_{P} > -x_{C} + (c + p)\big\}$ are the states already on the line $x_{P} = - x_{C} + (c + p)$. These states can be represented $(x_{C},x_{P}) = (c + \alpha, p - \alpha)$ for $\alpha > 0$. Note the $R_1$ transition rates of these states satisfy 
        $$ \lambda_{1}(c +\alpha, p -\alpha) = k_{1} (N - c - p) (M - \alpha - c) <  k_{1} (N - c - p) (M - c) = \lambda_{1}(c,p)$$
        If any state on the line $x_{P} = - x_{C} + (c+p)$ moves during the $R_1$ transition, then state $\mathcal{B}$ must necessarily undergo an $R_1$ transition also. $\mathcal{L}$ remains empty. \qed
        
        \item[$\boldsymbol{R_2}$:] (See above for effect of $R_2$ move on states in $\mathcal{T}$). States may move ``left" in the state space through an $R_2$ transitions, so we must check if a state will violate the assertion $\mathcal{L}$ is empty. If $\mathcal{B}$ is at $(c,p)$ when $k=n-1$, then a state $(x_{C},x_{P})$ for which $x_{P} \leq - x_{C} + (c +p)$ will move away from $\mathcal{L}$ during at $R_2$ transition, leaving $x_{P} = - x_{C} + (c + p)$ empty. Since $\mathcal{L}$ is empty, $R_2$ maps empty states to neighboring empty states and the vacated positions along this line are not filled. $\mathcal{L}$ remains empty. \qed
        
        \item[$\boldsymbol{R_3}$:] States may move ``up" in the state space through an $R_3$ transition. First check the assertion that $\mathcal{L}$ is empty. If $(x_{C},x_{P}) \in \mathcal{L}$, then if $x_{C} > c$ and $x_{P} < p$ an $R_3$ transition will map empty states to empty states. If $x_{C} > c$ but $x_{P} = p$, the $R_3$ transition will map points the states $(x_{C},p) \mapsto (x_{C}-1, p+1)$. But $\mathcal{T}$ was empty for $k=n-1$ and this move will map empty states in $\mathcal{L}$ to empty states in $\mathcal{T}$.
        
        Now check the assertion $\mathcal{T}$ is empty. All of the (possibly) nonempty candidates to move up into the triangle lie on the the line $x_{P} = p$. If $x_{C} > c$, then as noted above any states along this line are empty and the move will map empty states in $\mathcal{L}$ to nearby empty states in $\mathcal{T}$. If $x_{C} < c$, then the $R_3$ transition rate for such moves are $\lambda_{3}(x_{C},x_{P}) = k_{3} \, x_{C} < k_{3} \, c = \lambda_{3}(c,p)$. Therefore, $\mathcal{B}$ must necessarily undergo an $R_3$ move as well. $\mathcal{T}$ and $\mathcal{L}$ remain empty. \qed

        \item[$\boldsymbol{R_4}$:] States may move ``down" in the state space through an $R_4$ transition. The states on the line $x_{P} = p+1$ are empty for $x_{C} > c$, so at most an $R_4$ transition will map these empty states in $\mathcal{T}$ to empty states in $\mathcal{L}$. If $x_{C} < c$ then an $R_4$ transition will move (possibly) nonempty states away from $\mathcal{T}$. We must check the assertion that $\mathcal{L}$ is empty. The candidates states $(x_{C},x_{P})$ to move into $\mathcal{L}$ lie on the line $x_{P} = - x_{C} + (c + p)$ and can be represented with $(c + \alpha, p - \alpha)$ where $\alpha > 0$, but the $R_4$ transition will map these points diagonally along this line and they will not enter $\mathcal{L}$. $\mathcal{T}$ and $\mathcal{L}$ remain empty. \qed
    \end{itemize} 
\end{itemize}


\vspace{0.5in}
\subsection*{Proof of the Coupling Theorem}

\begin{theorem}{}
    Let $\mathbb{S}$ be the state-space for the Reversible Michaelis-Menten chemical reaction network and let the coordinate pair $(x_{c},x_{P})$ represent a state in $\mathbb{S}$. Define $\mathcal{A}$ to be the state initially found at $(0,0)$ and $\mathcal{B}$ to be the state initially found at $(0,N)$. Evolve the reaction network forward according to reactions $R_{i}$ ($i = 1,2,3,4$). If $\mathcal{A} = \mathcal{B} = (x_{c}^{\star},x_{P}^{\star})$ after an arbitrary number of reaction steps, then $(x_{c}^{\star},x_{P}^{\star})$ is the only non-empty state in $\mathbb{S}$.
\end{theorem}

    By contradiction. Suppose that $\mathcal{A} = \mathcal{B} = (x_{C}^{\star},x_{P}^{\star})$ when $k=n$ and suppose there is a nonempty state $(c,p)$ but that $(c,p) \neq (x_{C}^{\star},x_{P}^{\star})$. From Claim A, the set
    \begin{align*}
        \psi_{A} &=
            \big\{ x_{P} < x_{P}^{\star} \big\} 
            \cup \big\{ x_{C} < x_{C}^{\star}, \ x_{P} \leq - x_{C} + (x_{C}^{\star} + x_{P}^{\star}) \big\}
    \end{align*}
    is necessarily empty. From Claim B, the set
    \begin{align*}
        \psi_{B} &=
            \big\{ x_{P} > x_{P}^{\star} \big\} 
            \cup \big\{ x_{C} > x_{C}^{\star}, \ x_{P} > -x_{C} + (x_{C}^{\star} + x_{P}^{\star}) \big\}
    \end{align*}
    must also be empty. So the set of necessarily empty states in $\mathbb{S}$ when $\mathcal{A} = \mathcal{B}$ is
    \begin{align*}
        \psi &= \psi_{A} \cup \psi_{B} \equiv \mathbb{S}\backslash (x_{C}^{\star},x_{P}^{\star})
    \end{align*}
    So $(c,p) = (x_{C}^{\star},x_{P}^{\star})$. \qed


\bibliographystyle{vancouver}

\bibliography{masterbib.bib}

\end{document}